\documentclass[ amsmath,amssymb, prl, floatfix]{revtex4-2}
\usepackage[utf8]{inputenc}

\usepackage{graphicx}
\usepackage{dcolumn}
\usepackage{bm}
\usepackage{xcolor}
\usepackage{tikz}
\usepackage{pgfplots}
\usetikzlibrary{shapes}
\usetikzlibrary{calc}
\usetikzlibrary{arrows.meta}
\usepgfplotslibrary{fillbetween}
\usetikzlibrary{fadings}
\pgfplotsset{compat=1.17}

\usepackage{verbatim}

\usepackage{natbib}
\usepackage{url}

\begin{document}

\title{Magnetic signature of vertically migrating aggregations in the ocean}

\preprint{Preprint}

\date{\today}

\author{Matt K. Fu}
\affiliation{%
Graduate Aerospace Laboratories, California Institute of Technology, Pasadena, CA 91125
}%

\author{John O. Dabiri}
 \email{jodabiri@caltech.edu}
\affiliation{%
Graduate Aerospace Laboratories and Department of Mechanical and Civil Engineering, California Institute of Technology, Pasadena, CA 91125
}%


\begin{abstract}
The transport of heat and solutes by vertically migrating aggregations of plankton has long been explored as a potentially important source of ocean mixing \cite{Dewar2006,Katija2009,Houghton2018}. However, direct evidence of enhanced mixing due to these migrations remains challenging to obtain and inconclusive \cite{Kunze2019}. These shortcomings are due to the limitations of current measurement techniques, i.e., velocimetry techniques, which require \textit{a priori} knowledge of the precise aggregation location \cite{FernandezCastro2022} and typically trigger animal avoidance behavior from introducing instrumentation into the migration \cite{BeboitBird2016}. Here we develop a new approach to overcome these longstanding limitations by leveraging advancements in modern magnetometry to detect the flow-induced magnetic fields that naturally arise from seawater as it moves through the Earth's geomagnetic field  \cite{Tyler2003}. We derive quantitative predictions showing that these flow-induced magnetic fields in the vicinity of migrating aggregations have a strength proportional to the integrated fluid transport due to the migration. Importantly these magnetic signatures are potentially detectable remotely at a significant distance far from the aggregation and region of moving fluid with emerging quantum-enhanced magnetometry techniques such as Nitrogen-Vacancy centers in diamond \cite{Barry2020}. These results provide a new, testable framework for quantifying the significance of fluid transport in the ocean due to swimming organisms that may finally resolve a scientific debate \cite{Visser2007} with potentially enormous implications for our understanding of ocean dynamics and climate change. 
\end{abstract}

\maketitle

Biologically generated mixing from vertically migrating aggregations of plankton remains a poorly understood mechanism by which heat and solutes are potentially mixed in the ocean \citep{Wilhelmus2014,Dewar2006,Dabiri2009,Houghton2018}. While the induced flow and mixing associated with an isolated animal occur at the scale of the swimmer \cite{Visser2007}, plankton often exist in dense swarms over tens of meters in height and hundreds of meters in width \cite{Huntley2004,Sato2013}, and collectively traverse hundreds of meters during their diel vertical migrations \cite{Sato2013,Wiebe1979}. The emergence of aggregation-scale mixing eddies comparable to the stratification length scales of the water column has been proposed as a potential mechanism through which vertically migrating aggregations can induce appreciable mixing of the water column through which the organisms migrate \citep{Kunze2006,Dabiri2009}. Though recent laboratory studies provide evidence for such a mechanism \citep{Wilhelmus2014,Houghton2018,Houghton2019a}, direct environmental measurements of enhanced mixing due to vertically migrating aggregations in lakes and the ocean have proven less conclusive and challenging to obtain \citep{Rousseau2010,Lorke2010,Noss2014,Simoncelli2018,Kunze2019}. These challenges are due in large part to the practical difficulties associated with predicting, identifying, and quantifying instances of enhanced biomixing in the environment \cite{FernandezCastro2022}, especially given the spatiotemporal patchiness of their occurrence in the ocean. Moreover, \textit{in situ} measurements of local flow field are challenged by animal avoidance of instrumentation inserted in the water column in their vicinity \cite{BeboitBird2016}.

Magnetometry has emerged as a promising alternative to traditional velocimetry techniques \cite{Tyler1997,Tyler2006} to quantify large-scale marine flows, including vessel wakes \cite{Zou1998}, tsunami detection and parameterization \cite{Lin2021,Minami2021,Zhang2014}, wave measurements \cite{Podney1975,Davis1991}, and ocean current profiling \cite{Longuet-Higgins1954,Filloux1973,Lilley2004}. Instead of measuring the velocity field directly, these magnetic techniques instead measure the flow-induced magnetic fields that naturally arise when electrically conductive fluids, such as seawater, move through a magnetic field, such as the Earth's geomagnetic field \cite{Faraday1832,Podney1975,Tyler1997,Tyler2003}. In contrast to traditional velocimetry approaches, which measure localized quantities such as fluid parcel displacement, the flow-induced magnetic field is an inherently nonlocal feature related to integrated properties of the fluid flow. Importantly, flow-induced magnetic fields can potentially be detected remotely at a distance from the region of moving fluid. 

Recent simulations have suggested that turbulence generated by vertically migrating aggregations should also have a small, yet detectable magnetic signature \cite{Dean2019}. While the flow-induced magnetic signatures are typically several orders of magnitude smaller than the Earth's geomagnetic field strength, measurement of such signals is increasingly feasible due to rapid advances in the sensitivity, resolution, and availability of modern magnetometry techniques, especially quantum magnetometry techniques \cite{Dang2010,Wolf2015}. 

In this Letter, a new approach is proposed to overcome the limitations of conventional velocimetry techniques in quantifying vertical transport due to migrating aggregations via their distinct magnetic signatures. By scaling the electromagnetic field equations, the leading order dynamics that govern the magnetic perturbation created by a vertically migrating aggregation are derived and found to depend on the induced velocity field through a Poisson equation. Using this relationship, two representative models for the biologically induced velocity field are analyzed to predict the behavior of the corresponding flow-induced magnetic field. The first model is representative of high aspect ratio aggregation, such as those encountered in laboratory experiments \cite{Wilhelmus2014,Houghton2018,Houghton2019a,Fu2021}. In contrast, the second emulates the wider, low aspect ratio configurations observed in the field \cite{Sato2013,Wiebe1979}. Hence, magnetic detection of the migrations can potentially be accomplished without \textit{a priori} knowledge of the precise location of the aggregation and without triggering animal avoidance from the introduction of measurement instruments into the migration. 

Across both models, common features of the magnetic signature are observed. In the presence of a horizontal geomagnetic field such as that found near the equator, each of these velocity field models generates a magnetic signature, $\mathbf{b}$, that is poloidal, and whose vertical component has a strength proportional to the magnetic Reynolds number of the flow induced by the migration. Furthermore, the strength of this component is found to persist away from the aggregation and decay at a rate far slower than that of the corresponding velocity signature. Importantly, the magnetic signatures are predicted to be $\mathcal{O}(10-100\;\mathrm{pT})$, even at distances far removed from the aggregation, which are potentially detectable with modern and emerging magnetometry techniques.

\textit{Magnetic theory.}--- The motion of an electrically conducting fluid, such as seawater, through a magnetic field creates a corresponding electromagnetic signature. The electric current density, $\mathbf{j}$, induced by the motion of seawater can be determined from the version of Ohm's Law given by

\begin{equation}
    \mathbf{j} = \sigma\left(\mathbf{E} + \mathbf{u} \times \mathbf{B_{geo}}\right),
    \label{eq:ohms_law}
\end{equation}

\noindent where $\sigma$ is the electrical conductivity of the seawater (3-6 S/m), $\mathbf{E}$ is any applied or induced electric field, $\mathbf{u}$ is the fluid velocity field, and $\mathbf{B_{geo}}$ is the Earth's geomagnetic magnetic field (25,000-50,000 nT).  The resulting electric current, $\mathbf{j}$, creates a magnetic field perturbation, $\mathbf{b}$, which can be determined from the magnetostatic version of Ampere's Law as

\begin{equation}
    \nabla \times \mathbf{b} = \mu_0\; \mathbf{j}.
    \label{eq:amperes_law}
\end{equation}
Here, $\mu_0$ denotes the magnetic permeability of seawater is taken to be equal to the magnetic permeability of free space ($\mu_0 = 4\pi \times 10^{-7}\;\mathrm{H/m}$).

When temporal variations in the geomagnetic field are assumed to be small compared to temporal variations in the magnetic perturbation (i.e., $\partial \mathbf{B_{geo}}/\partial t \ll \partial \mathbf{b}/\partial t$), $\mathbf{E}$ in Eq. (\ref{eq:ohms_law}) can be related to the motionally-induced magnetic field perturbation, $\mathbf{b}$, through the Maxwell–Faraday Law of Induction:

\begin{equation}
    \frac{\partial \mathbf{b}}{\partial t} = - \nabla \times \mathbf{E}.
    \label{eq:faradays_law}
\end{equation}

\noindent From the incompressibility of the fluid flow and Gauss's Law of Magnetism, the velocity field and magnetic fields, respectively, are solenoidal (i.e., divergence free), following:

\begin{align}
\nabla \cdot \mathbf{b} = 0,\\
\nabla \cdot \mathbf{B_{geo}} = 0, \\
\nabla \cdot \mathbf{u} = 0.
\label{eq:gauss_law}
\end{align}

\noindent  Assuming $\sigma$ to be constant over the domain of interest and combining Eqs. (\ref{eq:ohms_law}) - (\ref{eq:gauss_law}) gives the relation

\begin{equation}
    \frac{\partial \mathbf{b}}{\partial t} =  \frac{1}{\mu_0\;\sigma} \nabla^2 \mathbf{b}  + \left(\mathbf{B_{geo}} \cdot \nabla \right) \mathbf{u} -  \left(\mathbf{u} \cdot \nabla \right) \mathbf{B_{geo}}.
    \label{eq:b_field}
\end{equation}

 Further simplification can be obtained by considering information related to the flows of interest. The leading order dynamics that determine $\mathbf{b}$ can be identified by replacing the variables in Eq. (\ref{eq:b_field}) with dimensionless variables scaled by a characteristic dimensional prefactor. The magnitude of each prefactor is representative of the relevant oceanic context and flow of interest, i.e., electrical conductivity ($\sigma = 5$ S/m), magnetic permeability ($\mu_0=4\pi\times10^{-7}$ H/m), length scale ($L=100$ m), and time scale ($T=1 $ hr).  The dimensionless variables are denoted with an $\sim$ overline and given by $\tilde{t}  = t/T$, $\mathbf{\tilde{b}} = \mathbf{b}/\beta$, $\mathbf{\tilde{r}} = \mathbf{r}/L = [x,y,z]/L =[\tilde{x},\tilde{y},\tilde{z}]$, and $\mathbf{u} = U \mathbf{\tilde{u}}$, where $\beta$ is magnetic field scale to be determined from the equations. Substituting these variables into Eq. (\ref{eq:b_field}) gives

\begin{equation}
  \left[\frac{L\beta}{U T \left|\mathbf{B_{geo}}\right|}\right] \frac{\partial \mathbf{\tilde{b}}}{\partial \tilde{t}}  =  \left[\frac{\beta}{\mu_0\;\sigma L U \left|\mathbf{B_{geo}}\right|}\right] \tilde{\nabla}^{2} \mathbf{\tilde{b}}  + \\ \left(\mathbf{\hat{B}_{geo}} \cdot \tilde{\nabla} \right) \mathbf{\tilde{u}} -   \left[\frac{ \delta B_{geo}}{\left|\mathbf{B_{geo}}\right|}\right] \left(\mathbf{\tilde{u}} \cdot \tilde{\nabla}  \right) \mathbf{\hat{B}_{geo}}  
    \label{eq:b_field_nd}.
\end{equation}

\noindent where $\mathbf{\hat{B}_{geo}}$ is the unit vector aligned with the geomagnetic field and $\delta B_{geo}$ is the scale of the variations in geomagnetic field strength over the domain size, $L$. In this formulation, all dimensionless variables are outside the brackets and are of order unity if appropriately scaled. The corresponding prefactors contained within the brackets denote the scale of each term in Eq. (\ref{eq:b_field_nd}) and quantify their relative importance to the dynamics. Assessing the magnitude of each scaling factor reveals that the first and last terms in Eq. (\ref{eq:b_field_nd}) are negligible for the flow of interest (see Supplementary Materials \cite{FuSuppMat2022} for further details), such that the leading order dynamics in Eq. (\ref{eq:b_field}) are governed by

\begin{equation}
    \nabla^2 \mathbf{b}  = -\mu_0\;\sigma\left(\mathbf{B_{geo}} \cdot \nabla \right) \mathbf{u}.
    \label{eq:b_field2}
\end{equation}
\noindent Furthermore, based on the scale of the leading order terms, the magnetic field perturbation scale, $\beta$, is found to scale as $\beta \sim \mathrm{R_m} \left|\mathbf{B_{geo}}\right|$, where  $\mathrm{R_m}=\mu_0 \sigma U L$ is the magnetic Reynolds number of the induced flow. The resulting 3D Poisson equation (\ref{eq:b_field2}) can be solved using a free space Green's function through the integral relation

\begin{equation}
     \mathbf{b} (\mathbf{r}) =  \iiint_V \frac{- \mu_0\sigma\, \left(\mathbf{B_{geo}} \cdot \nabla \right) \mathbf{u}(\mathbf{r'}) }{4 \pi \left| \mathbf{r} - \mathbf{r'} \right|}\, \mathrm{d}^3 r'.
    \label{eq:greens_func}
\end{equation}

Using the relationship in Eq. (\ref{eq:greens_func}), the magnetic signature, $\mathbf{b}$, can be determined from a given velocity field, $\mathbf{u}$, induced by an aggregation migrating through a geomagnetic field, $\mathbf{B_{geo}}$. 

\textit{Models for the induced flow field}---To determine magnetic signatures produced by vertical migrations, representative velocity fields are modeled for high and low aspect ratio configurations of migrating aggregations. The velocity field model for high aspect ratio aggregations is a unidirectional flow in the vertical ($z$) direction with Gaussian distribution in the horizontal plane (i.e., $xy$-plane). This type of induced velocity field is representative of those observed in laboratory experiments involving induced vertical migrations of zooplankton aggregations \citep{Wilhelmus2014,Houghton2018,Fu2021}. For an aggregation centered on the domain origin, the induced velocity field is given by

\begin{equation}
     \mathbf{u} (\mathbf{r}) = \left[u,\,v,\,w\right]= \left[0,\;0,\;W\, \textrm{exp}\left({\frac{-\left(x^2+y^2\right)}{2\varsigma_0^2}}\right)\right],
    \label{eq:velocity_field}
\end{equation}

\noindent where $W$ is the centerline vertical velocity, and $\varsigma_0$ is the characteristic finite width of the jet. Here, $x$ and $y$ are aligned with the geographic East-West and North-South directions, respectively, and $z$ is aligned with the vertical. In this high aspect ratio model, the induced flow is confined to a narrow radial extent in the horizontal plane relative to the size of the domain and has a homogeneous velocity signature along the vertical extent of the domain. 

The second velocity field model represents the effects of the migrating aggregation as a thin, thrust-generating disk (i.e., an actuator disk) \cite{Rankine1865} that is vertically moving at a steady climb rate. This modeling approach has been successfully applied to represent the induced fields of rotors \cite{Johnson1980}, wind turbines \cite{Shapiro2018}, and, more recently, krill \cite{Murphy2013} aggregations. In this case, the velocity field due to the migration is no longer homogeneous in the vertical direction but instead has a linearly expanding jet with a Gaussian velocity distribution in the horizontal direction. This expanding jet extends downstream from the aggregation position but minimally impacts the upstream fluid. In the frame of the migrating aggregation, the surrounding vertical velocity field for an aggregation centered on the domain origin climbing with an upward velocity, $W_v$, is given by

\begin{equation}
    w(x,y,z) = - W_v - \Delta w(z)\frac{D^2}{8\varsigma_0^2}\exp \left(\frac{-(x^2+y^2)}{2\varsigma_0^2 d_w(z)^2}\right),
    \label{eq:gauss_wake}
\end{equation}

\noindent where $\Delta w(z)$ is the vertical velocity surplus in the negative z (i.e. downward) direction along the jet centerline, $D$ is the nominal width of the aggregation, $\varsigma_0$ is the characteristic jet width where $\varsigma_0 = 0.235D$ \cite{Shapiro2018}, and $d_w(z)$ is the dimensionless spreading function of the jet as a function of distance downstream of the aggregation. The jet spreading function, $d_w(z)$, is modeled as a linear expansion similar to the Jensen wake model \cite{Jensen1983} and is given by the function
 
\begin{equation}
d_w(z) = 1+k_w \ln\left(1+\exp{\left(\frac{2(z-1)}{D}\right)}\right)
        \label{eq:jensen}
\end{equation}

\noindent with jet expansion coefficient set to $k_w\approx0.08$ as from \citet{Shapiro2018}, but can be adjusted without loss of generality. The corresponding centerline velocity surplus, $\Delta w$, that conserves vertical momentum in the aggregation jet is given by

\begin{equation}
    \Delta w(z) =\frac{\Delta w_0}{d_w^2(z)}\frac{1}{2}\left[1+\mathrm{erf}\left(\frac{z\sqrt{2}}{D}\right)\right]
        \label{eq:centerline_vel}
\end{equation}

 \noindent where $\Delta w_0$ denotes the induced velocity at the center of the aggregation position. When available, estimates for $\Delta w_0$ can be obtained from \textit{in situ} measurements \citep{Cisewski2009,Cisewski2021,Omand2021} or estimated \textit{a priori} from animal and aggregation parameters following the methodology of \citet{Houghton2019} (see Eq. 38 in Supplementary Materials\cite{FuSuppMat2022}). 

In each case, the geomagnetic field is taken to be constant over the domain of interest without declination ($B_x$) and inclination ($B_z$) such that $\mathbf{B} (\mathbf{r}) = \left[0,B_y ,\,0\right]$. The same analysis can be applied to other locations using the methods developed presently.

\textit{Structure of the magnetic signature}---Consider first the limit of Eq. (\ref{eq:velocity_field}) where $\varsigma_0\rightarrow{0}$ and $W\rightarrow{\infty}$ such that the vertical volume flux, $Q$, is finite, i.e., $Q=2\pi \varsigma_0^2 W=\textrm{constant}$. In this limit, the Gaussian velocity distribution simplifies to $\mathbf{u} (\mathbf{r}) = \left[u,\,v,\,w\right]= \left[0,\;0,\;Q\delta(x)\delta(y)\right]$ where $\delta$ is the Dirac delta distribution. This simplification allows Eq. (\ref{eq:greens_func}) to be solved analytically as

\begin{equation}
   \mathbf{b}= [b_x,b_y,b_z] = \\ \left[0,0,\frac{y\;H\;B_y\;\mu_0\;\sigma\;Q}{2\pi \left(x^2+y^2\right)\sqrt{H^2+x^2+y^2}}\right], 
    \label{eq:bz_dirac}
\end{equation}

\noindent where $2H$ is the vertical height over which the velocity field is integrated. From the relative directions of the geomagnetic field and fluid velocity, the magnetic perturbation manifests as a vertical magnetic field, $b_z$. This component decays inversely with the distance from the velocity signature near the aggregation and the inverse square of the distance in the far field. Furthermore, the magnitude of $b_z$ varies sinusoidally about the axis of the migration. While the above solution in Eq. (\ref{eq:bz_dirac}) is specific to the Dirac delta limit, it will be shown to generalize behavior derived from Eq. (\ref{eq:velocity_field}) throughout most of the domain.

\begin{figure}
    \centering
    \input{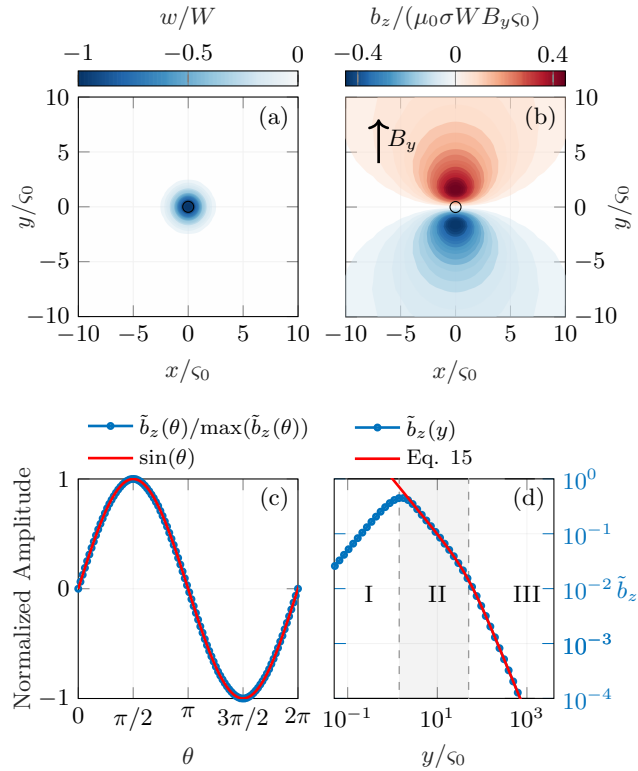}
    \caption{Dimensionless velocity and magnetic fields for high aspect ratio aggregation model. (a) Contours of dimensionless vertical velocity ($w/W$) in the $xy$-plane. (b) Contours of dimensionless vertical magnetic field component ($\tilde{b}_z = b_z B_y^{-1} \,\mathrm{R_m}^{-1}$) in the $xy$-plane. (c) Normalized strength of $\tilde{b}_z$ at a distance $\varsigma_0$ from the aggregation center vs. azimuthal position ($\theta$) in the $xy$-plane relative to the positive $x$ axis. (d) Variation of $\tilde{b}_z$  with distance along the $y$-axis (North-South direction). In (a) and (b), the black circle indicates the region of radius, $\varsigma_0$, centered on the aggregation location. In (c) and (d), blue circles show the vertical magnetic field signature computed numerically from Eq. (\ref{eq:greens_func}) and red lines show the limiting behavior for Dirac delta distribution from Eq. (\ref{eq:bz_dirac}). Scaling regimes in (d) are highlighted by different shading. Region I: $\tilde{b}_z \sim y/\varsigma_0$ for $y/\varsigma_0 < 1$, Region II: $\tilde{b}_z\sim \varsigma_0/y$ for $y/\varsigma_0 > 1$, and Region III: $\tilde{b}_z\sim \varsigma_0^2/y^2$ for $y/\varsigma_0 > H/\varsigma_0$. Here, $H = 50\varsigma_0$.}
    \label{fig1}
    \label{fig2}
\end{figure}

A contour map of the vertical velocity distribution is shown over the $xy$-plane in Fig. {\ref{fig1}(a)}. The corresponding results for the dimensionless vertical magnetic field, $\tilde{b}_z = b_z/(B_y\,\mathrm{R_m})$, are computed by numerically integrating Eq. (\ref{eq:greens_func}) and shown in Fig. {\ref{fig1}(b)} in the $xy$-plane, where $B_y$ is the strength of the North-South geomagnetic field component and, $\mathrm{R_m}=\mu_0 \sigma W \varsigma_0$ is the magnetic Reynolds number of the induced flow. Compared to the Gaussian velocity signature given by Eq. \ref{eq:velocity_field} (Fig. {\ref{fig1}a)}, the magnetic signature (Fig. {\ref{fig1}b)} persists much further away from the location of the induced flow. Furthermore, $\tilde{b}_z$ is found to vary sinusoidally with azimuthal angle about the vertical axis, as shown in Fig \ref{fig2}(c). 

In contrast to the Dirac delta solution (Eq. \ref{eq:bz_dirac}), the resulting magnetic signature exhibits three distinct scaling regimes with horizontal distance, as shown in {Fig \ref{fig2}(d)}. Within the vicinity of the velocity signature (Region I in Fig. {\ref{fig2}d}, $y/\varsigma_0 < 1$), the strength of the magnetic signature exhibits linear growth with distance away from the migration axis due to collocation with the downwelling. Outside of the induced velocity field, the behavior of the magnetic perturbation exhibits good agreement with the analytic expression given by Eq. (\ref{eq:bz_dirac}) for the Dirac delta limit (see Fig. {\ref{fig2}d}). In the region immediately outside the velocity signature (Region II, $y/\varsigma_0 > 1$), the magnetic perturbation decays inversely with distance from the migration (i.e., $\varsigma_0/y$) until $y/\varsigma_0 \approx H/\varsigma_0$ where the signal begins to exhibit a stronger decay with the inverse square of the distance from the migration. 

\begin{figure}[ht!]
    \centering
  \input{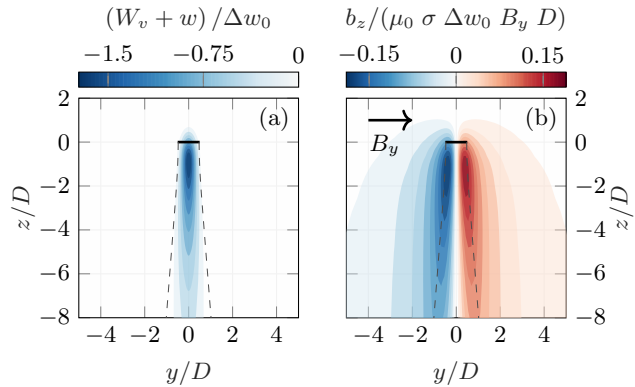}   
    \caption{Dimensionless velocity and magnetic fields for low aspect ratio aggregation model. In each panel, solid black lines represent actuator disk of diameter, $D$, and dashed lines indicates the nominal width of the aggregation jet given by $\pm2\varsigma_0d_w(z)$ in equation \ref{eq:jensen}. (a) Contour plot of vertical jet velocity generated by the aggregation in the $yz$-plane. (b) Contour plot of dimensionless vertical magnetic field component $(b_z/(\mu_0 \sigma \Delta w_0 B_y D))$ in the $yz$-plane. As in the previous models, the magnetic signature persists further from the aggregation location along the horizontal directions than the jet velocity signature. Here, $H = 400D$.}
    \label{fig3}
\end{figure}

Similar scaling behavior is observed using the velocity field model for low aspect ratio aggregations. The resulting jet velocity given by equations Eq. (\ref{eq:gauss_wake}) - (\ref{eq:centerline_vel}) is shown in {Fig. \ref{fig3}(a)} as a contour map in the $yz$-plane with a nominal jet spreading of $\pm2\varsigma d_w(z)$ (see Eq. \ref{eq:jensen}) shown as a dashed line. The corresponding induced magnetic field is shown in {Fig. \ref{fig3}(b)} as a contour map in the $yz$-plane with the same jet spreading function. Despite the reduced vertical extent of the induced flow, the magnetic signature is still observed to persist at horizontal distances much larger than the jet width for all values of $z$ downstream of the aggregation. Furthermore, the linear spreading of the velocity jet is associated with a commensurate gradual spreading of the magnetic perturbation downstream of the aggregation location. 

\begin{figure}[htpb]
    \centering
    \input{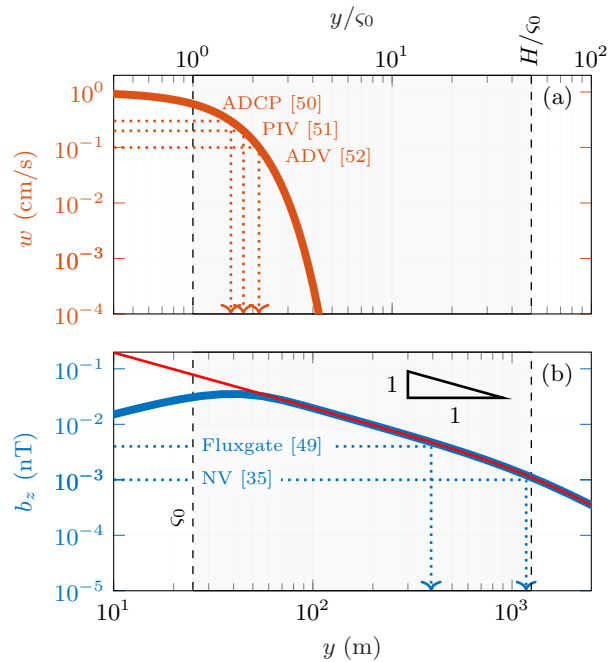}
    \caption{Profiles of representative dimensional (a) aggregation induced jet velocities and (b) corresponding vertical magnetic field strengths for the high aspect ratio model as a function of distance along the $y$-axis, i.e., North-South. Solid orange line in (a) shows the velocity field signature along the y-axis ($x=0$). Solid blue line in (b) shows the magnetic field signature along the y-axis ($x=0$), and the solid red line represents Eq. (\ref{eq:bz_dirac}). Dashed lines in orange and blue indicate the relative sensitivity limits of different velocimetry and magnetometry techniques, respectively. Triangle in (b) indicates $y^{-1}$ slope.} 
    \label{fig4}
\end{figure}

\textit{Detectability of biogenic signatures}---To assess the feasibility of detecting these magnetic signatures, representative values for each physical parameter are chosen and substituted for the dimensionless variables. Using $B_{geo}=25\mu$T \cite{WMM2022}, $\varsigma_0=25$ m \cite{Sato2013}, $\sigma = 5$ S/m, and $W=2$ $\mathrm{cm\cdot\mathrm{s}^{-1}}$ \citep{Cisewski2009,Cisewski2021,Omand2021}, the nominal scale of the vertical magnetic signature for the high aspect ratio model gives $\mathrm{R_m}B_y=\mu_0 \sigma W \varsigma_0  B_y  = 79$ pT. Recasting the data from the high aspect ratio model in terms of these parameters gives the distributions shown in Figs. \ref{fig4}(a) and \ref{fig4}(b) for both the vertical velocity and magnetic components as a function of distance along the $y$-axis. Superimposed on each distribution are the respective resolution/sensitivity limits for select measurement techniques for each parameter (see Supplementary Tables 1 and 2 \cite{FuSuppMatTable2022} for a detailed tabulation of velocimetry and magnetometry techniques, respectively). 

Common techniques for measuring velocity in the ocean such as Acoustic Doppler Current Profilers (ACDPs) \cite{NortekVector300,NortekAquedopp6000m,Park2021}, Acoustic Doppler Velocimeters (ADVs) \cite{NortekSignature1000,TeledyneWorkhorse2009,Cisewski2009,TeledyneOceanSurveyor2009,Cisewski2021}, and Particle Image Velocimetry (PIV) \cite{Bertuccioli1999,Katija2008,WangPIV2012,Jin2019} all have resolutions larger than 1 mm/s. Consequently, these techniques are suitable for observing upwelling and downwelling currents from migrating aggregates of zooplankton, which are typically on the order of a few centimeters per second \cite{Wilhelmus2014,Houghton2018,Cisewski2009,Cisewski2021,Omand2021}. However, as can be seen in Fig. \ref{fig4}(a), the Gaussian decay of the induced flow with distance from the migrating aggregation (Eq. \ref{eq:velocity_field}) confines the usefulness of these techniques to the immediate vicinity of the velocity signature, with each technique reaching its sensitivity floor within a distance of $2\varsigma_0-3\varsigma_0$ of the aggregation center. Quantifying the bulk fluid transport due to the migration with these velocimetry techniques is conceptually straightforward and involves measuring the vertical velocity distribution within the aggregation core and spatially integrating the results. However, in order to locate an instance of biogenic upwelling and downwelling via one of the localized velocimetry techniques (e.g., ADV), one would effectively need to be collocated with the aggregation, requiring \textit{a priori} knowledge of its precise location and potentially trigger avoidance behaviors by the animals. This limitation is not as severe for ADCPs, which are capable of measuring linear velocity profiles over significant ranges, though it is still necessary for the interrogation volume to intersect with the flow induced by the aggregation in order to detect the biogenic flow.

In contrast, the magnetic field signature has the advantage of being detectable at distances far removed from the velocity jet, at distances of up to a kilometer away. This feature is enabled by the slow spatial decay of the magnetic signature coupled with the advancements in the capability of modern vector magnetometry techniques. For example, commercial fluxgate magnetometers \cite{Bartington2022,Magson2S,THM1186} and emerging quantum sensing techniques such as Nitrogen-vacancy (NV) centers \cite{Wolf2015} have sensitivities on the order of $1-10 \, \mathrm{pT}/\sqrt{\mathrm{Hz}}$, which are theoretically able to detect this magnetic signature approximately an order of magnitude further away along the North-South axis. Though absolute magnetometers are often much more sensitive than their vector counterparts, the vertical alignment of the magnetic perturbation relative to the horizontal field might preclude the use of such techniques in the scenario.

\begin{figure}[ht!]
    \centering
    \input{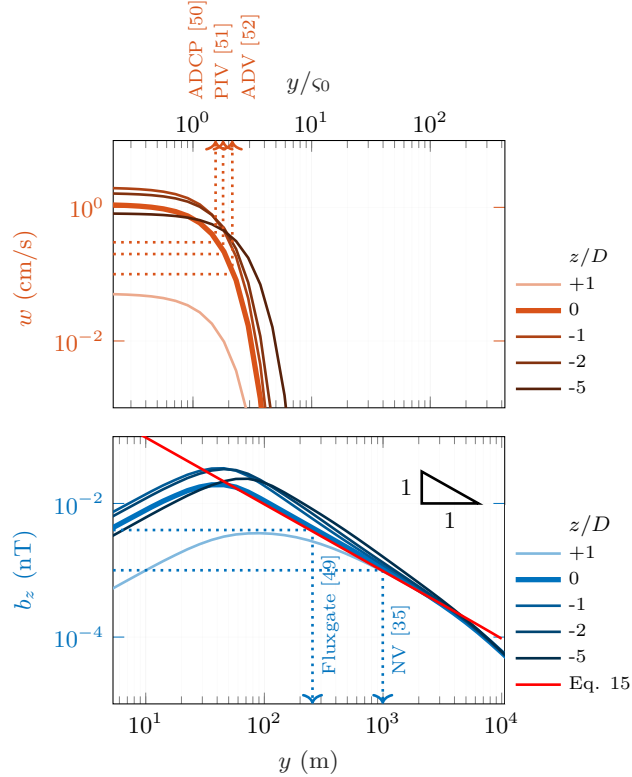}
    \caption{Profile of a representative dimensional (a) aggregation jet velocities and (b) corresponding vertical magnetic field strengths for the low aspect ratio model as a function of distance along the $y$-axis at different heights. Solid orange and blue lines show the vertical jet velocity and vertical magnetic field component, respectively, along the y-axis ($x=0$) at vertical locations $z/D=-5,-2,-1,0,\;\mathrm{and}\;1$. Darker shades of each respective color indicate lower heights, with the thick line denoting the height of the aggregation itself.  Dotted lines indicate the typical resolution or sensitivity limit of corresponding velocimetry and magnetometry techniques. The red line in (b) indicates the nominally equivalent magnetic signature generated by Eq. \ref{eq:bz_dirac}. Here, $H = 400D$}
    \label{fig5}
\end{figure}

Similar behavior is observed in the analogous results from the low aspect ratio model, shown in Fig. (\ref{fig5}). As in Fig. (\ref{fig4}), select profiles of the vertical magnetic and velocity components are shown along the $y$-axis in {Fig. \ref{fig5}(a)} at various vertical locations 
with the respective resolution/sensitivity limits of different measurement techniques. Substituting the same parameters into these distributions ($B_{geo}=25\mu$T, $D=106$ m, $\sigma = 5$ S/m, and $\Delta w_0=1$ cm/s) again gives a magnetic signature scale of $\mu_0 \sigma \Delta w_0 B_y D= 166$ pT.  Similar to the previous analysis, common techniques such as Acoustic Doppler Current Profilers (ACDPs) \cite{NortekVector300,NortekAquedopp6000m,Park2021} and Acoustic Doppler Velocimeters  \cite{NortekSignature1000,TeledyneWorkhorse2009,Cisewski2009,TeledyneOceanSurveyor2009,Cisewski2021} are all still suitable for observing upwelling and downwelling currents from migrating aggregates of zooplankton even up to $5D$ downstream of the aggregation. 
As shown in {Fig. \ref{fig5}(a)}, the Gaussian decay of the velocity signature still confines the usefulness of velocimetry techniques to the immediate vicinity of the jet. However, because of the gradual expansion of the jet downstream of the aggregation, the horizontal distance from the axis of the migration where the velocity signature can be detected gradually increases downstream of the migration. By comparison, the decay of the magnetic signatures at vertical locations downstream of the aggregation (see {Fig. \ref{fig5}b)} is slightly faster than the $y^{-1}$ predicted by the other models. Interestingly, for a given horizontal location, there is also a relative enhancement of the magnetic signature with downstream distance from the migration outside the velocity jet due to the entrainment and jet spreading. Despite these differences, the magnetic field profile collocated with the vertical plane of the aggregation appears in good agreement with the high aspect ratio model given in Eq. \ref{eq:bz_dirac} and shown by the red line. 

The persistence of this inverse decay of the induced magnetic field with distance from the aggregation across these different models facilitates a concise relationship through which the magnetic perturbation can be related to the biogenic upwelling and downwelling. For magnetic field measurements obtained in the $y^{-1}$ regime (Region II in Fig. \ref{fig2}d),  $\tilde{b}_z=b_z/(B_y \mu_0 \sigma W \varsigma_0)=\varsigma_0\,y/(x^2+y^2)$. Rearranging these terms reveals a new relationship for the volumetric flow rate driven by the migrating aggregation where

\begin{equation}
    Q = 2\pi W \varsigma_0^2 = \frac{b_z }{B_y \mu_0 \sigma}\frac{2\pi(x^2+y^2)}{y}.
\label{eq:flowrate}
\end{equation}
In the above relationship, all properties of the aggregation and jet are contained on the left-hand side in the form of the volumetric flow rate. This quantity can be theoretically determined directly from suitable measurements of the magnetic perturbation, $b_z$, and their position relative to the aggregation (i.e., $x$ and $y$) provided that the relevant environmental properties (i.e., $B_y$ and $\sigma$) are known. {Though a single measurement of $b_z$ is theoretically sufficient, practical implementations may require mapping $b_z$ over at multiple positions around the aggregation. Symmetric deployment of several magnetometers allows for the rejection of common sources of magnetic noise (e.g., the ionosphere) and verification of the $y^{-1}$ behavior. Similarly, these measurements can be phase-locked with a corresponding measurement of acoustic back-scattering, velocity, and photosynthetically available radiation \citep{Omand2021} to correlate the resulting magnetic signal to the migration behavior of the aggregation. By complementing traditional tools with this new magnetic approach, it may finally be possible to quantify the significance of fluid transport and mixing in the ocean due to migrating aggregations of zooplankton.}

This work was supported by the U.S. National Science Foundation (NSF) Alan T. Waterman Award.

\bibliographystyle{apsrev4-2}
\bibliography{references.bib}

\end{document}


\title{Supplementary Materials: Magnetic signature of vertically migrating aggregations in the ocean}
\preprint{Preprint}

\date{\today}

\author{Matt K. Fu}
\affiliation{%
Graduate Aerospace Laboratories, California Institute of Technology, Pasadena, CA 91125
}%

\author{John O. Dabiri}
 \email{jodabiri@caltech.edu}
\affiliation{%
Graduate Aerospace Laboratories and Department of Mechanical and Civil Engineering, California Institute of Technology, Pasadena, CA 91125
}%

\maketitle

\section{Magnetic Equations of Motion in the Ocean}
\label{sec:theory}

The electric current density, $\mathbf{j}$, induced by the motion of seawater can be determined from Ohm's Law given by

\begin{equation}
    \mathbf{j} = \sigma\left(\mathbf{E} + \mathbf{u} \times \mathbf{B_{geo}}\right),
    \label{eq:ohms_law}
\end{equation}

\noindent where $\sigma$ is the electrical conductivity of the seawater (3-6 S/m), $\mathbf{E}$ is any applied or induced electric field, $\mathbf{u}$ is the fluid velocity field, and $\mathbf{B_{geo}}$ is the Earth's geomagnetic magnetic field (25,000-50,000 nT).  The resulting electric current, $\mathbf{j}$, in turn, has an associated magnetic field perturbation, $\mathbf{b}$, which can be determined from the non-relativistic (magnetostatic) version of Ampere's Law as
\begin{equation}
    \nabla \times \mathbf{b} = \mu_0\; \mathbf{j}.
    \label{eq:amperes_law}
\end{equation}
Here, $\mu_0$ denotes the magnetic permeability of seawater ($\mu_0 = 4\pi \times 10^{-7}\;\mathrm{H/m}$), which is taken to be equal to the magnetic permeability of free space.

Substituting equation \ref{eq:amperes_law} into \ref{eq:ohms_law} for $\mathbf{j}$ gives the relation

\begin{equation}
    \mathbf{E}  = \frac{\nabla \times \mathbf{b}}{\mu_0\;\sigma}  - \mathbf{u} \times \mathbf{B_{geo}}.
    \label{eq:e_field}
\end{equation}

 If the temporal variations in the geomagnetic field are assumed to be small compared to temporal variations in the magnetic perturbation (i.e., $\partial B_{geo}/\partial t \ll \partial b/\partial t$), then the electric field in equation \ref{eq:e_field} can be related to the motionally-induced magnetic field, $\mathbf{b}$, through the Maxwell–Faraday Law of Induction:

\begin{equation}
    \frac{\partial \mathbf{b}}{\partial t} = - \nabla \times \mathbf{E}.
    \label{eq:faradays_law}
\end{equation}

Taking the curl of equation \ref{eq:e_field} allows equation \ref{eq:faradays_law} to be expressed as  

\begin{equation}
    \frac{\partial \mathbf{b}}{\partial t} = - \nabla \times \left( \frac{\nabla \times \mathbf{b}}{\mu_0\;\sigma}  - \mathbf{u} \times \mathbf{B_{geo}} \right).
    \label{eq:b_field}
\end{equation}

\noindent The first and second terms on the right-hand side of equation \ref{eq:b_field} can be expanded and rewritten using the vector identities

\begin{equation}
 -\nabla \times \left( \frac{\nabla \times \mathbf{b}}{\mu_0\;\sigma} \right)= -\frac{ \nabla \times \left(\nabla \times \mathbf{b} \right)}{\mu_0\;\sigma} - \nabla\left(\frac{1}{\mu_0\sigma}\right)\times\left(\nabla\times\mathbf{b}\right) = - \frac{1}{\mu_0\;\sigma} \left( \nabla \left(\nabla \cdot \mathbf{b}  \right) - \nabla^2 \mathbf{b}  \right)   - \nabla\left(\frac{1}{\mu_0\sigma}\right)\times\left(\nabla\times\mathbf{b}\right),
    \label{eq:curlofacurl}
\end{equation}

\noindent and  

\begin{equation}
\nabla \times \left( \mathbf{u} \times \mathbf{B_{geo}}\right) = \mathbf{u}  \left(\nabla \cdot  \mathbf{B_{geo}} \right) -  \mathbf{B_{geo}}\left( \nabla \cdot \mathbf{u} \right) + \left(\mathbf{B_{geo}} \cdot \nabla \right) \mathbf{u} -  \left(\mathbf{u} \cdot \nabla \right) \mathbf{B_{geo}}, 
    \label{eq:crossofacurl}
\end{equation}
respectively. Using equations \ref{eq:curlofacurl} and \ref{eq:crossofacurl}, equation \ref{eq:b_field} can be expressed as
\begin{equation} 
    \frac{\partial \mathbf{b}}{\partial t} = - \frac{\left( \nabla \left(\nabla \cdot \mathbf{b}  \right) - \nabla^2 \mathbf{b}  \right)}{\mu_0\;\sigma}  - \nabla\left(\frac{1}{\mu_0\sigma}\right)\times\left(\nabla\times\mathbf{b}\right)  +
    \mathbf{u}  \left(\nabla \cdot  \mathbf{B_{geo}} \right) -  \mathbf{B_{geo}}\left( \nabla \cdot \mathbf{u} \right) + \left(\mathbf{B_{geo}} \cdot \nabla \right) \mathbf{u} -  \left(\mathbf{u} \cdot \nabla \right) \mathbf{B_{geo}}.
    \label{eq:b_field2}
\end{equation}

\noindent Because the fluid flow is assumed to be incompressible, the velocity field will be solenoidal (i.e., divergence free), following:

\begin{equation}
\nabla \cdot \mathbf{u} = 0.
\label{eq:incompressible}
\end{equation}

Similarly, by Gauss' Law of Magnetism, both magnetic fields are also solenoidal:
\begin{equation}
\nabla \cdot \mathbf{b} = 0 \quad ; \quad \nabla \cdot \mathbf{B_{geo}} = 0.
\label{eq:gauss_law}
\end{equation}

Using the constraints from equations \ref{eq:incompressible} and \ref{eq:gauss_law}, equation \ref{eq:b_field2} reduces to

\begin{equation}
    \frac{\partial \mathbf{b}}{\partial t} =  \frac{1}{\mu_0\;\sigma} \nabla^2 \mathbf{b} - \nabla\left(\frac{1}{\mu_0\sigma}\right)\times\left(\nabla\times\mathbf{b}\right)  + \left(\mathbf{B_{geo}} \cdot \nabla \right) \mathbf{u} -  \left(\mathbf{u} \cdot \nabla \right) \mathbf{B_{geo}}.
    \label{eq:b_field3}
\end{equation}

To further simplify the relation between seawater motion, $\mathbf{u}$, and the induced magnetic field perturbation, $\mathbf{b}$, additional information related to the flows of interest can be considered. The leading order dynamics of the magnetic field perturbation, $\mathbf{b}$, can be identified by substituting the variables in equation \ref{eq:b_field3} for dimensionless variables that have been scaled by an appropriate, dimensional prefactor. These dimensionless variables are denoted with an $\sim$ overline and given by

\begin{align*}
 \tilde{t}& = t/T,\\
\mathbf{\tilde{b}} & = \mathbf{b}/\beta,  \\
\mathbf{\tilde{r}} = \mathbf{r}/L &= [x,y,z]/L =[\tilde{x},\tilde{y},\tilde{z}],\\
\mathbf{u} &= U \mathbf{\tilde{u}},\\
\mathbf{\sigma} &= \sigma_0 \mathbf{\tilde{\sigma}}.\\
\end{align*}

\noindent The magnitude of each prefactor (e.g., $L$, $U$, $\sigma_0$, and $T$) is determined by the flow configuration of interest. The corresponding magnetic field perturbation scale, $\beta$, remains to be determined. Substituting these variables into equation \ref{eq:b_field3} gives

\begin{equation}
  \left[\frac{\beta}{T}\right] \frac{\partial \mathbf{\tilde{b}}}{\partial \tilde{t}}  =  \left[\frac{\beta}{\mu_0\;\sigma_0 L^2}\right] \left(\tilde{\nabla}^{2} \mathbf{\tilde{b}} - \tilde{\nabla}\left(\frac{1}{\tilde{\sigma}}\right)\times\left(\tilde{\nabla}\times\mathbf{\tilde{b}}\right)\right)  + \left[\frac{U \left|\mathbf{B_{geo}}\right|}{L}\right]\left(\mathbf{\hat{B}_{geo}} \cdot \tilde{\nabla} \right) \mathbf{\tilde{u}} -  \left[\frac{U\;\delta B_{geo}}{L}\right] \left(\mathbf{\tilde{u}} \cdot \tilde{\nabla}  \right) \mathbf{\hat{B}_{geo}},  
    \label{eq:b_field4a}
\end{equation}

\noindent where $\mathbf{\hat{B}_{geo}}$ is the unit vector aligned with the direction of the geomagnetic field and $\delta B_{geo}$ is the scale of the variations in the geomagnetic field strength over the domain of interest. This choice of scaling takes a conservative approach where the gradients in the velocity, conductivity, and magnetic perturbation fields are assumed to scale the same as the scaling prefactor divided by the length scale, $L$. In this formulation, all the dimensionless variables are outside the brackets and are assumed to be on the order of unity if appropriately scaled. The prefactors contained within the brackets denote the scale of each term in the equation, quantifying their relative importance to the dynamics. Normalizing equation \ref{eq:b_field4a} by the scale $\left[U\;\left|\mathbf{B_{geo}}\right|L^{-1}\right]$ gives

\begin{equation}
  \left[\frac{L\beta}{U T \left|\mathbf{B_{geo}}\right|}\right] \frac{\partial \mathbf{\tilde{b}}}{\partial \tilde{t}}  =  \left[\frac{\beta}{\mu_0\;\sigma_0 L U \left|\mathbf{B_{geo}}\right|}\right] \left(\tilde{\nabla}^{2} \mathbf{\tilde{b}}  - \tilde{\nabla}\left(\frac{1}{\tilde{\sigma}}\right)\times\left(\tilde{\nabla}\times\mathbf{\tilde{b}}\right) \right) + \left(\mathbf{\hat{B}_{geo}} \cdot \tilde{\nabla} \right) \mathbf{\tilde{u}} -  \left[\frac{ \delta B_{geo}}{\left|\mathbf{B_{geo}}\right|}\right] \left(\mathbf{\tilde{u}} \cdot \tilde{\nabla}  \right) \mathbf{\hat{B}_{geo}}  
    \label{eq:b_field4}
\end{equation}

\noindent such that each of the prefactors is now a dimensionless quantity, and the scale of the $\left(\mathbf{\hat{B}_{geo}} \cdot \tilde{\nabla} \right) \mathbf{\tilde{u}}$ term is normalized to unity. To identify which terms in equation \ref{eq:b_field4} are of leading order, the relative magnitudes of the terms involving the magnetic field perturbation, $\mathbf{b}$, can be compared. The ratio between the prefactors of the unsteadiness term on the left-hand side and the Laplacian and conductivity gradient terms on the right-hand side is given by

\begin{equation*}
 \left. \left[\frac{L\beta}{U T \left|\mathbf{B_{geo}}\right|}\right] \middle/  \left[\frac{\beta}{\mu_0\;\sigma_0 L U \left|\mathbf{B_{geo}}\right|}\right] = \left[\frac{L^2\mu_0\;\sigma_0 }{T}\right] \right. .
    \label{eq:dbdt}
\end{equation*}

Assigning representative values for the scaling parameters based on the relevant oceanic context of $\sigma_0 = 6$ S/m, $\mu_0=4\pi\times10^{-7}$ H/m, $L=1000$ m, and $T=1 $ hr, yields ${L^2\mu_0\sigma_0 T^{-1}}=\mathcal{O}(10^{-3})$. The value of this ratio indicates that contributions from the unsteadiness term to the dynamics of equation \ref{eq:b_field4} are approximately three orders of magnitude smaller than those of the Laplacian and conductivity gradient terms and can likely be neglected. In this case, it can also be concluded that the scaling prefactor of the Laplacian term in equation \ref{eq:b_field4}, $\left[{\beta(\mu_0\;\sigma_0 L U \left|\mathbf{B_{geo}}\right|)^{-1}}\right]$, must be of leading order to ensure that the magnetic perturbation, $\mathbf{b}$, is still included in the leading order dynamics. 

Finally, it remains to be established if the final term in equation \ref{eq:b_field4}, which scales as $\left[\delta B_{geo}\left|\mathbf{B_{geo}}\right|^{-1}\right]$, is of leading order. The geomagnetic field strength, $\left|\mathbf{B_{geo}}\right|$, is found to vary less than $0.1\%$ over domain sizes of $L=$1000 m \cite{WMM2022} indicating that $\left[\delta B_{geo}\left|\mathbf{B_{geo}}\right|^{-1}\right]=\mathcal{O}(10^{-3})$. This similarly small ratio indicates that contributions from the last term are also not of leading order for the flows of present interest and can similarly be neglected. Having now assessed the scale of each prefactor, the leading order dynamics in equation \ref{eq:b_field4} are found to be order unity, and it can be concluded that $\left[\beta^{-1}\mu_0\;\sigma_0 L U \left|\mathbf{B_{geo}}\right|\right] =\mathcal{O}(1)$ to ensure it is retained in the leading order dynamics. Following the above scaling analysis, the leading order terms from equation \ref{eq:b_field4} that relate the seawater motion, $\mathbf{u}$, to the induced magnetic field perturbation, $\mathbf{b}$, are given by:

\begin{equation}
 0= \tilde{\nabla}^{2} \mathbf{\tilde{b}}  -  \tilde{\nabla}\left(\frac{1}{\tilde{\sigma}}\right)\times\left(\tilde{\nabla}\times\mathbf{\tilde{b}}\right) + \left(\mathbf{\hat{B}_{geo}} \cdot \tilde{\nabla} \right) \mathbf{\tilde{u}}, 
    \label{eq:b_field5}
\end{equation}
or in dimensional terms,
\begin{equation}
       0 =  \frac{1}{\mu_0\;\sigma} \nabla^2 \mathbf{b}  + \left(\mathbf{B_{geo}} \cdot \nabla \right) \mathbf{u} - \nabla\left(\frac{1}{\mu_0\sigma}\right)\times\left(\nabla\times\mathbf{b}\right).
    \label{eq:laplace_eq}
\end{equation}

In cases where the electrical conductivity of the seawater is assumed to be horizontally homogeneous (i.e., $\sigma=\sigma(z)$), further simplification of equation \ref{eq:laplace_eq} can be observed when expressed in component wise form as:

\begin{subequations}
\begin{align}
   \left(\frac{\partial^2 b_x}{\partial x^2}+\frac{\partial^2 b_x}{\partial y^2}+\frac{\partial^2 b_x}{\partial z^2}\right)  &= -\mu_0\sigma(z)\left(B_x \frac{\partial u}{\partial x} + B_y \frac{\partial u}{\partial y}+ B_z \frac{\partial u}{\partial z}\right)+\frac{1}{\sigma(z)}\frac{\partial\sigma(z)}{\partial z}\left(\frac{\partial b_z}{\partial x}-\frac{\partial b_x}{\partial z}\right)   \label{eq:b_field3_vecx} \\    
    \left(\frac{\partial^2 b_y}{\partial x^2}+\frac{\partial^2 b_y}{\partial y^2}+\frac{\partial^2 b_y}{\partial z^2}\right)  &=  -\mu_0\sigma(z)\left(B_x \frac{\partial v}{\partial x} + B_y \frac{\partial v}{\partial y}+ B_z \frac{\partial v}{\partial z}  \right) +\frac{1}{\sigma(z)}\frac{\partial\sigma(z)}{\partial z}\left(\frac{\partial b_z}{\partial y}-\frac{\partial b_y}{\partial z}\right) \label{eq:b_field3_vecy} \\
   \left(\frac{\partial^2 b_z}{\partial x^2}+\frac{\partial^2 b_z}{\partial y^2}+\frac{\partial^2 b_z}{\partial z^2}\right)  &=  -\mu_0\sigma(z)\left(B_x \frac{\partial w}{\partial x} + B_y \frac{\partial w}{\partial y}+ B_z \frac{\partial w}{\partial z}  \right) \label{eq:b_field3_vecz}
\end{align}
\end{subequations}

\noindent where $B_{geo} = [B_x,\,B_y,\,B_z]$. While the horizontal components of the magnetic perturbation, $b_x$ and $b_y$, each depend on the vertical gradient of electrical conductivity (see equations \ref{eq:b_field3_vecx} and \ref{eq:b_field3_vecy}), the equation for $b_z$ (equation \ref{eq:b_field3_vecz}) no longer has such dependencies when $\sigma = \sigma(z)$. Furthermore, because equation \ref{eq:b_field3_vecz} has no dependency on $b_x$ or $b_y$, it is a 3D Poisson equation for $b_z$ and can be solved using the free space Green's function as

\begin{equation}
 b_z (\mathbf{r}) =   \iiint_V \frac{- \mu_0\sigma(z)}{4 \pi \left| \mathbf{r} - \mathbf{r'} \right|}\, \left(B_x \frac{\partial w}{\partial x'} + B_y \frac{\partial w}{\partial y'}+ B_z \frac{\partial w}{\partial z'} \right) \mathrm{d}^3 r'
 \label{eq:greenz}
 \end{equation}
         
\noindent where the local conductivity field is included in the forcing term of the integrated function. The same is not true, however, for the horizontal components which depend on horizontal gradients in $b_z$. This complication requires alternative approaches such as determining the vertical component of the magnetic field, $b_z$, and then iteratively solving for the horizontal components or solving the entire set of equations numerically through a relaxation method.

In cases when $\sigma$ is constant over the entire domain, then equation \ref{eq:laplace_eq} simplifies instead to
\begin{equation}
       0 =  \frac{1}{\mu_0\;\sigma} \nabla^2 \mathbf{b}  + \left(\mathbf{B_{geo}} \cdot \nabla \right) \mathbf{u}
    \label{eq:laplace_eq2}
\end{equation}
which is a vectorized Poisson equation in 3D. Here, each component can be solved with the free space Green's function for the 3D Poisson equation through the integral relation

\begin{equation}
     \mathbf{b} (\mathbf{r}) =  \iiint_V \frac{- \mu_0\sigma\; \left(\mathbf{B_{geo}}(\mathbf{r'}) \cdot \nabla \right) \mathbf{u}(\mathbf{r'}) }{4 \pi \left| \mathbf{r} - \mathbf{r'} \right|}\, \mathrm{d}^3 r'.
     \label{eq:greens_func} 
\end{equation}

Expressing each vector component of equation \ref{eq:greens_func} with $B_{geo} = [B_x,\,B_y,\,B_z]$ gives 

\begin{subequations}
    \begin{align}
         b_x (\mathbf{r}) &=  \iiint_V \frac{- \mu_0\sigma\;}{4 \pi \left| \mathbf{r} - \mathbf{r'} \right|}\, \left(B_x \frac{\partial u}{\partial x'} + B_y \frac{\partial u}{\partial y'}+ B_z \frac{\partial u}{\partial z'} \right)\mathrm{d}^3 r' 
        \label{eq:greens_func_x} \\
         b_y (\mathbf{r}) &=  \iiint_V \frac{- \mu_0\sigma\;}{4 \pi \left| \mathbf{r} - \mathbf{r'} \right|}\, \left(B_x \frac{\partial v}{\partial x'} + B_y \frac{\partial v}{\partial y'}+ B_z \frac{\partial v}{\partial z'} \right) \mathrm{d}^3 r'
        \label{eq:greens_func_y} \\
         b_z (\mathbf{r}) &=   \iiint_V \frac{- \mu_0\sigma\;}{4 \pi \left| \mathbf{r} - \mathbf{r'} \right|}\, \left(B_x \frac{\partial w}{\partial x'} + B_y \frac{\partial w}{\partial y'}+ B_z \frac{\partial w}{\partial z'} \right) \mathrm{d}^3 r'
        \label{eq:greens_func_z} 
    \end{align}
\end{subequations}

\noindent where $\mathbf{r} = (x,y,z)$ and $\mathbf{r'} = (x',y',z')$. The above relations are valid inside the ocean and the Earth's surface. Above the ocean, where there is no electrical current or conductivity, the fields are determined by Laplace's equation for the scalar potential $\nabla V=-\mathbf{b}$ such that $\nabla^2 V = -\nabla \cdot \mathbf{b} = 0$.

\section{Magnetic signature of vertical flow induced by swimming aggregations}
\label{sec:models}
Having established the relationship between the seawater motion, $\mathbf{u}$, and the induced magnetic field perturbation, $\mathbf{b}$, the magnetohydrodynamic signature of vertically migrating aggregations can be derived for representative velocity fields. In the following section, three different models for the biologically generated velocity field will be considered. First, the flow induced from vertically migrating aggregations is modeled using a Dirac delta distribution in the $xy$-plane (i.e., horizontal plane) having a strength, $Q$, representing the volumetric flow rate caused by the aggregation wake. This configuration mimics the scenario in which the induced flow is confined to a narrow radial extent in the horizontal plane relative to the size of the domain and assumes a homogeneous velocity signature along the vertical extent of the domain.

The next model considers the induced flow to have a Gaussian distribution in the horizontal plane with a characteristic finite width, $\varsigma_0$, with centerline vertical velocity, $W$, along the vertical axis. Similar to the previous model, the velocity signature is assumed to extend uniformly along the vertical extent of the domain.

In the final model, the effect of the aggregation is modelled using an actuator disk \cite{Rankine1865} with a steady rate of vertical climb. In this case, the velocity field due to the migration is no longer homogeneous in the vertical direction. Instead, the velocity signature is modeled as linearly expanding wake with a Gaussian velocity profile that extends downstream from the aggregation position and has a negligible influence upstream.

When subjected to a horizontal geomagnetic field such as that present near the equator, each of these velocity field models generates a magnetic signature, $\mathbf{b}$, that is poloidal, i.e., having primarily a vertical component. The magnitude of the vertical component is found to decay inversely with distance from the induced flow and be modulated sinusoidally around the vertical axis of the migration direction. Importantly, the slower decay of the magnetic signature (i.e., $b\sim r^{-1}$) compared to the velocity signature (i.e., $w\sim e^{-r}$) indicates that the physical signature of a vertical migration of swimming plankton is potentially detectable with modern magnetometry techniques from distances where the induced flow cannot be detected. The magnetic field could provide additional insight into the bulk fluid transport associated with biologically induced flow, as well as the flow induced by other vertical transport processes in the ocean.

\subsection{Dirac Delta Migration Model}
\label{sec:models:dirac}
The biologically generated velocity field stemming from the vertically migrating aggregations, $\mathbf{u} (\mathbf{r})$, is modelled first with a Dirac delta distribution in the horizontal plane given by:
\begin{equation}
     \mathbf{u} (\mathbf{r}) = \left[u,\,v,\,w\right]= \left[0,\;0,\;\pm Q
     \delta(x)\delta(y)\right],
    \label{eq:dirac_delta}
\end{equation}
 where $\pm Q$ is the volumetric flow rate associated with the induced flow, respectively. This velocity distribution is most representative when the characteristic width of the aggregation is small compared to the vertical extent of the velocity signature and the distances at which the magnetic signature is being measured. 
 Similar to the previous section, the magnetic field signature can be related to the specific flow parameters through dimensional analysis. In this model, there are eight physical parameters: the magnitude of the magnetic signature, $b$, the geomagnetic field strength, $B_{geo}$, volumetric flow rate, $Q$, electrical conductivity of the fluid, $\sigma$, magnetic permeability of seawater, $\mu_0$, fluid density, $\rho$, kinematic viscosity, $\nu$, and distance from the migration, $\varrho$, along with four associated dimensions: mass, length, time, and electric current. Through the applications of the Buckingham-$\pi$ theorem, four dimensionless groups can be identified such that the functional dependence of $b$ can be expressed as 
\begin{equation}
    \frac{b}{B_{geo}} \cdot = f_1\left(   N ,   \mathrm{R_\varrho} , m\right)
\end{equation}
where $N \equiv B_{geo}^{2}\varrho^3\sigma/(\rho Q)$ is the Stuart Number, $\mathrm{R_m} \equiv  Q \varrho^{-1} \sigma \mu_0$ is the dimensionless distance from the aggregation, and $m \equiv \nu \cdot \sigma \cdot \mu_0$ is the magnetic viscosity ratio. For parameter values representative of the flow of present interest, (i.e., $B_{geo}=25\mu$T, $Q=40\,\mathrm{m}^3/\mathrm{s}$ m, $\sigma = 6$ S/m, $\rho=1.025$ g/mL, $U=5$ mm/s, and $\varrho = 1000$m), the Stuart Number is $N=O(10^{-5})$, indicating that electromagnetic forces within the fluid are much smaller than inertial forces, consistent with the assumptions of our model. Similarly, the dimensionless magnetic signature $b \cdot B_{geo}^{-1}$ is expected to exhibit a functional dependence on the dimensionless distance, $\mathrm{R_m} \equiv  Q \varrho^{-1} \sigma \mu_0$, which has the form of a magnetic Reynolds number. 

To specify the functional form of the magnetic signature, $\mathbf{b}$, the model velocity field can be analyzed using the Green's function approach described in section \ref{sec:theory} with a known geomagnetic field, $\mathbf{B_{geo}}$. Here, the geomagnetic field is taken to be constant over the domain without declination ($B_x$) and inclination ($B_z$) such that

\begin{equation}
     \mathbf{B} (\mathbf{r}) = \left[0,B_y ,\,0\right]
    \label{eq:B_geo}
\end{equation}

\noindent where $B_x$ is prescribed to be aligned with the East-West direction, $B_y$ is aligned with the geographic North-South direction and $B_z$ is aligned with the vertical. This choice of geomagnetic field is representative of equatorial regions, where biologically generated mixing has been proposed as a potential contributor to the Meridional Overturning Circulation (MOC).  \cite{Munk1966,Dewar2006,Dabiri2010}. It also follows from equations \ref{eq:greens_func_z} - \ref{eq:greens_func_z}
 that the homogeneity of the vertical velocity field along the vertical ($z$) direction restricts the dependence of the vertical component of the magnetic signature, $b_z$, to only horizontal components of the geomagnetic field (i.e., $B_x$ and $B_y$).

Substituting equations \ref{eq:dirac_delta} and \ref{eq:B_geo}  into equations \ref{eq:greens_func_x}-\ref{eq:greens_func_z} gives

\begin{equation}
     \mathbf{b}(\mathbf{r})=[b_x,\;b_y,\;b_z] =  \left[0,\;0,\; \iiint_V \frac{B_y\;\mu_0\;\sigma\;Q\delta(x')\frac{\textrm{d}\delta(y')}{\textrm{d}y'}}{4\pi \sqrt{(x-x')^2+(y-y')^2+(z-z')^2}}\, \mathrm{d}x' \mathrm{d}y' \mathrm{d}z'\right].
    \label{eq:bz_delta3}
\end{equation}

Using the identity 
\begin{equation}
    \int f(x)\delta'(x) \textrm{d}x = \int -f'(x)\delta(x) \textrm{d}x, 
\end{equation}
integration in the $x$ and $y$ directions yields
\begin{align*}
     b_z &=  B_y\;\mu_0\;\sigma\;Q \iiint_V \frac{1}{4\pi \sqrt{x^2+(y-y')^2+(z-z')^2}}\, \frac{\textrm{d}\delta(y')}{\textrm{d}y'}\mathrm{d}x' \mathrm{d}y' \mathrm{d}z'\\
     b_z &= B_y\;\mu_0\;\sigma\;Q \iint \frac{(y-y')}{4\pi \left(x^2+(y-y')^2+(z-z')^2\right)^{3/2}}\, \delta(y')\mathrm{d}y' \mathrm{d}z'\\
    b_z &=   B_y\;\mu_0\;\sigma\;Q \int \frac{y}{4\pi \left(x^2+y^2+(z-z')^2\right)^{3/2}}\,\mathrm{d}z'.\\
   b_z &= B_y\;\mu_0\;\sigma\;Q\frac{y\;H}{2\pi \left(x^2+y^2\right)\sqrt{H^2+x^2+y^2}}\\
\end{align*}
for constant $\sigma$. In scenarios when $\sigma$ varies along the $z$ direction, it should remain in the integrand. Finally, integrating in the $z$-direction from $-H$ to $H$, where $2H$ is the height of the velocity signature, yields
\begin{align*} 
    b_z &= B_y\;\mu_0\;\sigma\;Q\frac{y\;H}{2\pi \left(x^2+y^2\right)\sqrt{H^2+x^2+y^2}}.\\
\end{align*}
In the limit of $H\rightarrow\infty$, the relation for the magnetic field signature given by
\begin{equation}
    \frac{b_z}{B_y} = \frac{y\;\mu_0\;\sigma\;Q}{2\pi \left(x^2+y^2\right)} =  \frac{1}{2\pi}\frac{\mu_0\;\sigma\;Q\;\sin{\theta}}{\sqrt{x^2+y^2}}.
    \label{eq:bz_dirac}
\end{equation}

This resulting expression is consistent with the above dimensional analysis, revealing that the strength of the magnetic field perturbation decays inversely with distance from the velocity signature and is modulated sinusoidally by the azimuthal angle, $\theta$, from the positive $x$-axis.

\subsection{Gaussian Jet Model}
\label{sec:models:gauss}

The second model for $\mathbf{u} (\mathbf{r})$ is given by a unidirectional flow along the $z$ direction with Gaussian distribution along the $x$ and $y$ directions.  For distributions centered on the domain origin, the velocity field for induced flow can be expressed as
\begin{equation}
     \mathbf{u} (\mathbf{r'}) = \left[u,\,v,\,w\right]= \left[0,\;0,\;\pm W\, \textrm{exp}\left({\frac{-\left(x'^2+y'^2\right)}{2\varsigma_0^2}}\right)\right]
    \label{eq:velocity_field}
\end{equation}

\noindent where $W$ is the velocity scale of the induced flow and $\varsigma_0$ is the characteristic width of the jet. A contour map of the vertical velocity distribution for downwelling is shown over the $xy$-plane in figure \ref{fig:field}(a). Integrating equation \ref{eq:velocity_field} over the $xy$-plane reveals that the net volumetric flow rate associated with this induced flow model is $Q=\pm 2\pi\,W\,\varsigma_0^2$. Taking the limit of $\varsigma_0\rightarrow{0}$ and $W\rightarrow{\infty}$ where the volume flux, $Q$, is finite, i.e., $Q=2\pi \varsigma_0^2 W=\textrm{constant}$, recovers the Dirac delta distribution from section \ref{sec:models:dirac} given by equation \ref{eq:dirac_delta}.

Similar to the previous model, the Earth's magnetic field is taken as constant with a negligible declination ($B_x$) and inclination ($B_z$) and $B_y$ aligned with the North-South direction. 

With the velocity fields (equation \ref{eq:velocity_field}) and applied magnetic fields (equation \ref{eq:B_geo}) known, equations \ref{eq:greens_func_x}-\ref{eq:greens_func_z} can be simplified as

\begin{equation}
     \mathbf{b} (\mathbf{r}) = \left[b_x,\,b_y,\,b_z\right] \\= \left[0,\;0,\;\mu_0\;\sigma \iiint_V \frac{1}{4 \pi \left| \mathbf{r} - \mathbf{r'} \right|}\, \left(B_y \frac{\partial w}{\partial y'}\right)\mathrm{d}^3 r'\right]
    \label{eq:upwelling}
\end{equation}
where
\begin{equation}
\frac{\partial w}{\partial y'} = y'\frac{W }{\varsigma_0^2}\, \textrm{exp}\left({\frac{-\left(x'^2+y'^2\right)}{2\varsigma_0^2}}\right).
\label{eq:dwdy}
\end{equation}
Expanding the $b_z$ component gives
\begin{equation}
     b_z = \;\mu_0\;\sigma \iiint_V \frac{B_y}{4 \pi \left| \mathbf{r} - \mathbf{r'} \right|}\, \left( y'\frac{W }{\varsigma_0^2}\, \textrm{exp}\left({\frac{-\left(x'^2+y'^2\right)}{2\varsigma_0^2}}\right)\right)\mathrm{d}^3 r'.
    \label{eq:bz_1}
\end{equation}

Rescaling the equation with length scale, $\varsigma_0$, gives a pre-factor of $\varsigma_0^3$ and new dimensionless variables of integration: $\mathbf{\tilde{r}} = \mathbf{{r}}/\varsigma_0$, $\tilde{x}' = x'/\varsigma_0$, $\tilde{y}' = y'/\varsigma_0$, and $\tilde{z}' = z'/\varsigma_0$:

\begin{equation}
     \tilde{b}_z(\mathbf{\tilde{r}}) = \frac{b_z(\mathbf{\tilde{r}})}{\mu_0\;\sigma\;B_y\;W\;\varsigma_0} = \frac{1}{4 \pi} \iiint \frac{\tilde{y}'}{ \left| \mathbf{\tilde{r}} - \mathbf{\tilde{r}'} \right| }\, \textrm{exp}\left({\frac{-\left(\tilde{x}'^2+\tilde{y}'^2\right)}{2}}\right)\mathrm{d} \tilde{x}'\,\mathrm{d} \tilde{y}'\,\mathrm{d} \tilde{z}',
    \label{eq:bz_nd}
\end{equation}
which can be solved numerically.
\subsubsection{Numerical Approach and Results}
\label{sec:gauss:results}
To compute the integral in equation \ref{eq:bz_nd}, the velocity field is discretized onto a 3D domain ranging from $-5\varsigma_0$ to $5\varsigma_0$ in each of the $x$ and $y$ directions and from $-320\varsigma_0$ to $320\varsigma_0$ in the vertical (i.e., z-direction) using $[N_x,\,N_y,\,N_z] = [64,\,64,\,2560]$ gridpoints to mimic a long vertical extent of downwelling. The magnetic field was evaluated on a larger domain on the $xy$-plane (i.e., $z=0$) at 50 logarithmically spaced locations ranging from $x,y = 0.01 - 1000$. A grid convergence study of the velocity field discretization was conducted to verify that the root-mean-square differences for the different resolutions were less than $1\%$ of the global root-mean-square variations. Further, a domain study using different heights from $H= 80\varsigma_0-640\varsigma_0$ was conducted to observe their effects on the power law behavior ranges. All computations were performed on an Nvidia RTX Quadro 5000 GPU.

\begin{figure}[htpb]
    \centering
    \input{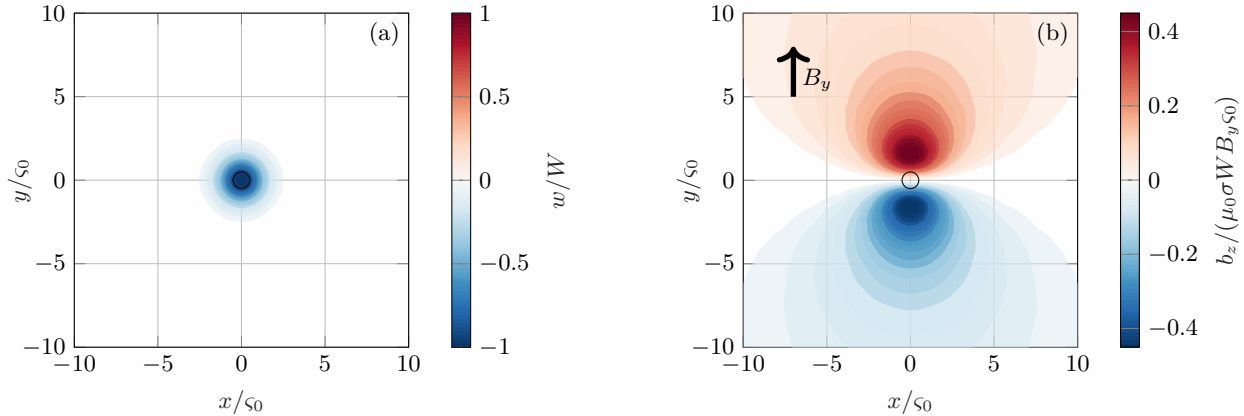}
    \caption{Plots of the dimensionless velocity and magnetic fields. (a) Contours of the vertical velocity component in the $xy$-plane. (b) Contours of the dimensionless vertical magnetic field component $(b_z/(\mu_0 \sigma W B_y \varsigma_0))$ in the $xy$-plane. The lobed structure of the magnetic field follow a sinusoidal dependence with azimuthal angle. The peak values occur approximately $\sqrt{2}\varsigma_0$ away from the center of the migration. In each panel, the black circle indicates the region of radius, $\varsigma_0$, centered on the aggregation location.  The Gaussian velocity distribution in (a) decays significantly faster with distance from the migration than the corresponding magnetic signature (b). }
    \label{fig:field}
\end{figure}

\begin{figure}[htpb]
    \centering
    \input{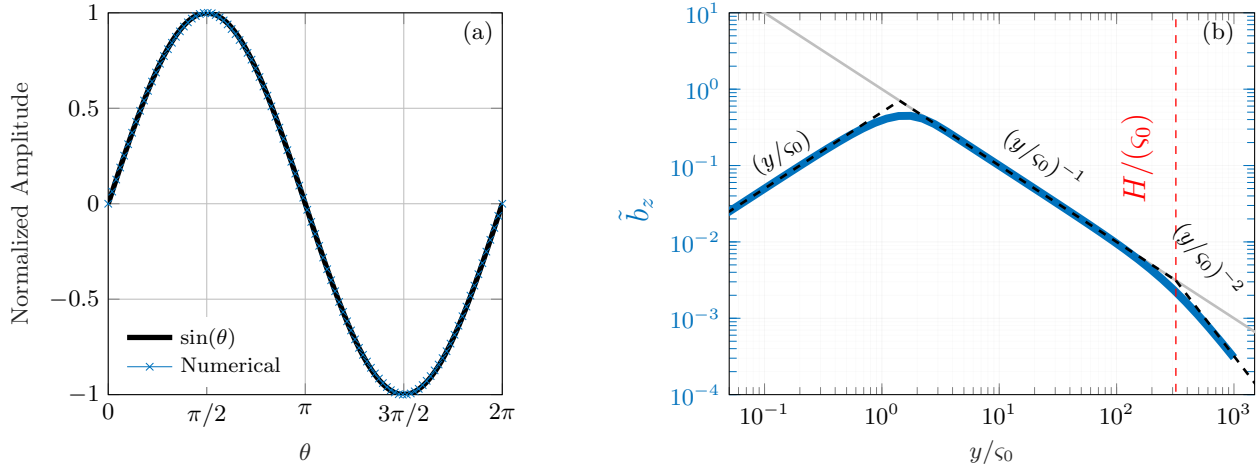}
    \caption{Comparison of numerical results with Rankine magnetic field model.  (a) Normalized magnetic field amplitude as a function of azimuthal position in the $xy$-plane relative around the center of the migration. The azimuthal angle is taken with respect to the positive $x$ axis. The amplitude variation exhibited by the magnetic field (blue x symbols) is computed for location at a distance $\varsigma_0$ from the aggregation center and found to agree well with the sinusoidal approximation (solid black line). (b) Decay of magnetic signature with distance along the $y$-axis (North-South direction). Blue lines show the vertical magnetic field signature computed numerically along the $y$-axis. Dashed black lines show the Rankine model results along the $y$-axis. Three distinct regimes are observed 1) a $y/\varsigma_0$ growth in the migration, 2) a $\varsigma_0/y$ decay for $y/\varsigma_0 >\sqrt{2}$, and , 3) a $(\varsigma_0/y)^2$ decay for $y/\varsigma_0 > H/\varsigma_0$.  Gray line shows the solution for the Dirac Delta velocity model. Here, $H = 320\varsigma_0$.}
    \label{fig:rankine}
\end{figure}

The numerical results for the dimensionless vertical magnetic field $\tilde{b}_z$ on the $xy$-plane are shown in figure \ref{fig:field}(b). Compared to the Gaussian velocity signature shown in figure \ref{fig:field}(a), the magnetic signature persists much further away from the location of induced flow and exhibits a lobed structure. More precisely, the variation of the magnetic signature with azimuthal angle follows a sinusoidal dependence within the $xy$-plane and is shown in figure \ref{fig:rankine}(a) as a function of azimuthal locations where $\varrho \equiv \sqrt{x^2+y^2} =\varsigma_0$, consistent with the simpler Dirac delta model.

The variation of the magnetic signature with distance is found to exhibit three distinct scaling regimes, which are shown in figure \ref{fig:rankine}(b) at locations along the $y$-axis. The first regime occurs within the vicinity of the velocity signature ($y/\varsigma_0 \ll1$). Here, the strength of the magnetic signature is largest due to the proximity to the finite velocity gradients and grows with distances away from the migration axis. The second regime begins outside the velocity signature region ($y/\varsigma_0 > 1$) and extends to $y/\varsigma_0 \approx H/\varsigma_0$. Here, the signal begins to decay inversely with distance from the migration (i.e., $(y/\varsigma_0)^{-1}$) and most closely emulates the behavior and assumptions of the Dirac delta model. The final regime is encountered at distances comparable to the height of the migration. There, the signal begins to exhibit a stronger decay, scaling nominally the inverse square of distance from the migration (i.e.,$(y/\varsigma_0)^{-2}$), and is dominated by the effects of the finite domain size. 

The collection of these distinct behaviors in the magnetic signature are qualitatively analogous to the Rankine model of a vortex in viscous flow. In that model, the azimuthal velocity magnitude is found to increase linearly within a viscously dominated core and decay inversely with distances outside of the core. Here, an analogous behavior is observed in the magnetic signature, albeit with an additional sinusoidal modulation along the azimuth. The analogous Rankine model for the magnetic signature is given by the piecewise equation:

 \begin{equation}
\tilde{b}_z(\mathbf{\tilde{r}})= \begin{cases} 
      y/(2\varsigma_0) & \varrho/\varsigma_0\leq \sqrt{2} \\
      \varsigma_0y/\varrho^2 & \sqrt{2}\leq \varrho/\varsigma_0\leq H/(\varsigma_0) \\
      H\varsigma_0y/\varrho^3 & \varrho/\varsigma_0 > H/(\varsigma_0)
      \label{eq:rankine}
  \end{cases}
  \end{equation}
 where $\varrho = \sqrt{x^2+y^2}$.

\begin{figure}[htpb]
    \centering
    \input{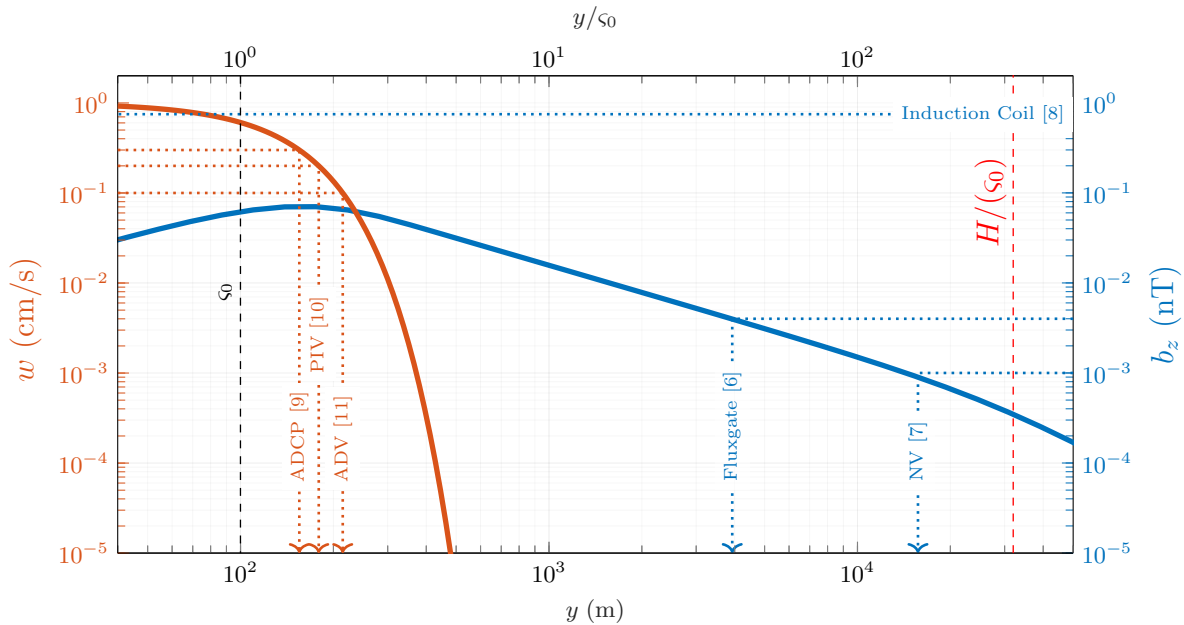}
    \caption{Profile of the dimensional magnetic field strength compared with the velocity signature along the $y$-axis. Solid blue lines show the magnetic field signature along the y-axis ($x=0$). Solid orange lines show the velocity field signature along the y-axis ($x=0$). Here we choose representative values of $B_{geo}=25\mu$T, $\varsigma_0=100$ m, $\sigma = 5$ S/m, $W=1$ cm/s, giving a representative magnetic signature magnitude of $157$ pT. Dashed lines in orange and blue indicate the relative sensitivity limits of different velocimetry and magnetometry techniques, respectively. Red dashed line represents the height of the resolved velocity wake.}
    \label{fig:profile}
\end{figure}

To determine the measurement sensitivity required to detect the magnetic signature, representative values for each parameter are chosen as $B_{geo}=25\mu$T, $\varsigma_0=100$ m, $\sigma = 5$ S/m, $W=1$ cm/s, and substituted for the dimensionless variables. The nominal scale of the vertical magnetic signature is found to be  $\mu_0 \sigma B_y W \varsigma_0 = 157$ pT. Recasting the data from figure \ref{fig:rankine}(b) in terms of these dimensional parameters gives the distributions shown in figure \ref{fig:profile} for both the vertical magnetic and velocity components as a function of distance from the aggregation center. Superimposed on each distribution are the resolution or sensitivity limit for select measurement techniques for each parameter. A detailed tabulation of velocimetry and magnetometry techniques is compiled in Tables \ref{tab:tab1_vel_params} and \ref{tab:tab2_mag_params}, respectively. For the velocity field, common techniques such as Acoustic Doppler Current Profilers (ACDPs) \cite{NortekVector300,NortekAquedopp6000m,Park2021}, Acoustic Doppler Velocimeters \cite{NortekSignature1000,TeledyneWorkhorse2009,Cisewski2009,TeledyneOceanSurveyor2009,Cisewski2021}, and Particle Image Velocimetry (PIV) \cite{Bertuccioli1999,Katija2008,WangPIV2012,Jin2019}, all have resolution limits close to a few millimeters per second. Consequently, these techniques are suitable for observing upwelling and downwelling currents from migrating aggregates of zooplankton, which are typically on the order of a few centimeters per second \cite{Wilhelmus2014,Houghton2018,Cisewski2009,Cisewski2021,Omand2021}. However, as can be seen in figure \ref{fig:profile}, the Gaussian decay of the velocity signature confines the usefulness of these techniques to the immediate vicinity of the velocity signature, with each technique reaching its sensitivity floor within a distance of $2\varsigma_0-3\varsigma_0$ of the aggregation center. Quantifying the bulk fluid transport due to the migration with these velocimetry techniques is conceptually straightforward and involves measuring the vertical velocity distribution within the aggregation core and spatially integrating the results. 

Even though the resolvable velocity signature is confined to a range comparable to the aggregation width, $\varsigma_0$, the magnetic signature is potentially detectable at ranges at least an order of magnitude larger. This feature is enabled both by the persistence of the magnetic signature due to the inherent nonlocality of the magnetic field and the advancements in the capability of modern magnetometry techniques. For example, commercial fluxgate magnetometers \cite{Bartington2022,Magson2S,THM1186} and emerging quantum sensing techniques such as Nitrogen-vacancy (NV) centers \cite{Wolf2015} have sensitivities on the order of $1-10 \, \mathrm{pT}/\sqrt{\mathrm{Hz}}$, which are theoretically detect this magnetic signature up to $100\varsigma_0$ away along the N-S axis.

A potential benefit of this feature is the ability to locate instance of biogenic mixing via their magnetic signature. As seen in figure \ref{fig:profile}, in order to locate an instance of biogenic upwelling and downwelling via one of the localized velocimetry techniques (e.g., ADV), one would effectively need to be collocated with the aggregation wake. This limitation is not as applicable to ADCPs, which are capable of measuring linear velocity profiles at-a-distance. However, it is still necessary for the interrogation volume to intersect with the aggregation velocity wake in order to detect the biogenic flow. In contrast, the magnetic signature is inherently a nonlocal quantity that extends far beyond that of the velocity wake. While the inherent scale of the biogenic magnetic signature is small compared to the Earth's geomagnetic field, such a signal is potentially detectable with existing magnetometry techniques, including commercially available fluxgate magnetometer. Determining the size of the magnetic signature and mapping its distribution can potentially be accomplished with even a handful of magnetometers while also helping to identify the location of the aggregation and its velocity signature.

\subsection{Low Aspect Ratio Aggregations - Actuator Disk Model}
\label{sec:models:disc}
\subsubsection{Induced velocity due to vertically migrating aggregations}
In contrast to the previous models, where the velocity signature was assumed to have a long extent in the vertical direction, the current model considers the case where the vertical extent of the aggregation, $H$, is not only finite but much smaller than its characteristic width of the aggregation, $D$. In this low aspect ratio configuration, the aggregation can be modeled as an actuator disk \cite{Rankine1865,Houghton2019} in a steady rate of vertical climb. The velocity signature due to the migration is no longer homogeneous in the vertical direction but is instead confined to the region downstream of the aggregation location. In this section, the velocity induced by the vertical migration is related to the properties of the animals that comprise the aggregation following \citet{Houghton2019}. This induced velocity is then synthesized with the linearly expanding wake model based on actuator disk theory.

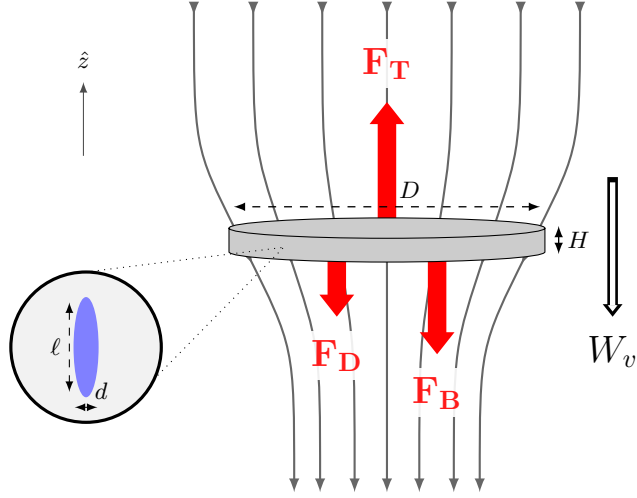
\begin{figure}[ht!]
\centering
\begin{tikzpicture}[>=latex,shorten >=2pt,shorten <=2pt,shape aspect=3,my triangle/.style={-{Triangle[width=\the\dimexpr1.8\pgflinewidth,length=\the\dimexpr1.2\pgflinewidth]}}]

    \newcommand\dsp{0.1cm}
    \newcommand\R{2cm}
    \newcommand\xx{\R/3};
    \newcommand\yy{\R/3*2};
    \definecolor{mycolor1}{rgb}{0.00000,0.44700,0.74100}%
    \definecolor{mycolor2}{rgb}{0.85000,0.32500,0.09800}%
    \definecolor{strellor}{rgb}{0.4,0.4,0.4}%

    \draw[strellor, thick,>->] (\R+\xx-\dsp, 2.5*\yy) .. controls (\R+\xx-\dsp,\yy)  .. (\R-\dsp, 0cm);
    \draw[strellor, thick,<-<] (\R-\xx-\dsp, -2.5*\yy) .. controls (\R-\xx-\dsp, -\yy)  .. (\R-\dsp, 0cm);
    \draw[strellor, thick,>->] (-\R-\xx+\dsp, 2.5*\yy) .. controls (-\R-\xx+\dsp, \yy)  .. (-\R+\dsp, 0cm);
    \draw[strellor, thick,<-<] (-\R+\xx+\dsp, -2.5*\yy) .. controls (-\R+\xx+\dsp, -\yy)  .. (-\R+\dsp, 0cm);
    
    \newcommand\pr{1.5};
    \draw[strellor, thick,>->] (\R/\pr+\xx/\pr, 2.5*\yy) .. controls (\R/\pr+\xx/\pr,\yy)  .. (\R/\pr, 0cm);
    \draw[strellor, thick,<-<] (\R/\pr-\xx/\pr, -2.5*\yy) .. controls (\R/\pr-\xx/\pr, -\yy)  .. (\R/\pr, 0cm);
    \draw[strellor, thick,>->] (-\R/\pr-\xx/\pr, 2.5*\yy) .. controls (-\R/\pr-\xx/\pr,\yy)  .. (-\R/\pr, 0cm);
    \draw[strellor, thick,<-<] (-\R/\pr+\xx/\pr, -2.5*\yy) .. controls (-\R/\pr+\xx/\pr, -\yy)  .. (-\R/\pr, 0cm);

    \renewcommand\pr{3.1};
    \draw[strellor, thick,>->] (\R/\pr+\xx/\pr, 2.5*\yy) .. controls (\R/\pr+\xx/\pr,\yy)  .. (\R/\pr, 0cm);
    \draw[strellor, thick,<-<] (\R/\pr-\xx/\pr, -2.5*\yy) .. controls (\R/\pr-\xx/\pr, -\yy)  .. (\R/\pr, 0cm);
    \draw[strellor, thick,>->] (-\R/\pr-\xx/\pr, 2.5*\yy) .. controls (-\R/\pr-\xx/\pr,\yy)  .. (-\R/\pr, 0cm);
    \draw[strellor, thick,<-<] (-\R/\pr+\xx/\pr, -2.5*\yy) .. controls (-\R/\pr+\xx/\pr, -\yy)  .. (-\R/\pr, 0cm);
    
    \draw[strellor, thick,>->] (0,2.5*\yy) -- (0, -2.5*\yy);

    \draw[line width=7pt,my triangle,draw=red](-\R/3*0,2.5mm) -- (-\R/3*0, 2cm) node  [above, text = red, fill = white, opacity = 0.9] {\Large $\mathbf{F_T}$};
    \draw[line width=7pt,my triangle,draw=red](-\R/3,-1mm) -- (-\R/3, -1cm) node  [below,text =red,fill = white, opacity = 0.9] {\Large $\mathbf{F_D}$};
    \draw[line width=7pt,my triangle,draw=red](\R/3,-1mm) -- (\R/3, -1.5cm) node  [below,text = red,fill = white, opacity =0.9] {\Large $\mathbf{F_B}$};
    
    
\node[cylinder,draw=black, fill = black!20!white,aspect=1.25,minimum height=0.01cm,minimum width=2*\R+2*\dsp,shape border rotate=90] (A) {};

  
    \draw [black!70,->]    let \p1 = ($ (A.after bottom) - (A.before bottom) $),
        \n1 = {0.5*veclen(\x1,\y1)-\pgflinewidth},
        \p2 = ($ (A.bottom) - (A.after bottom)!.5!(A.before bottom) $),
        \n2 = {veclen(\x2,\y2)-\pgflinewidth}
  in ([xshift=-1pt-2*\R,yshift=-\n2-3.5pt+\R/2]A.top) -- ([xshift=-1pt-2*\R,yshift=-\n2+1cm+\R/2]A.top) node [above,black]  {$\hat{z}$};
  
    \draw [<->,line width =0.5pt,dashed]([yshift=6mm] A.before bottom) -- ([yshift=6mm] A.after bottom) node [midway, above,xshift=3mm] {$D$};
    \draw [<->,line width =0.5pt] ([xshift=5pt,yshift=-1mm] A.before bottom) -- ([xshift=5pt,yshift=1mm] A.after top) node [midway, right] {$H$};

    \draw[line width=4pt,my triangle,draw=black](1.5*\R,1cm) -- (1.5*\R,-1cm) node  [below,text = black] {\Large $W_v$};
    \draw[line width=2pt,my triangle,draw=white](1.5*\R,1cm-2pt) -- (1.5*\R,-1cm+2pt) ;

    \draw[black,dotted] (-\R/1.5, 0) -- (-2*\R, -\yy+\R/2);
    \draw[black,dotted] (-\R/1.5, 1pt) -- (-2*\R+\R/3, -\yy-\R/3);
    \filldraw[color=black, fill=black!5, very thick] (-2*\R, -\yy) circle (\R/2);
    \fill[color=black, fill=blue!50] (-2*\R, -\yy) ellipse (\R/12 and \R/3);
    \draw[black,dashed,<->] (-2*\R-\R/9, -\yy+\R/3) -- (-2*\R-\R/9, -\yy-\R/3) node [midway, left] {$\ell$};
    \draw[black,dashed,<->] (-2*\R-\R/9, -\yy-\R/2.5) -- (-2*\R+\R/9, -\yy-\R/2.5) node [midway, above right] {$d$};



\end{tikzpicture}
\caption{Diagram of the actuator disk model in its inertial frame.  The aggregation is modeled as a porous disk of diameter, $D$, and height, $H$, shown in gray. The aggregation is assumed to climb upwards along the positive $\hat{z}$ direction at a constant velocity, $W_v$. The force balance of the actuator disk is shown where the upwardly directed thrust force ($F_T$) shown in blue is balanced by the downward directed buoyancy ($F_B$) and drag ($F_D$) forces shown in orange. The individual animals are modeled as prolate spheroids, indicated in light blue with body length, $\ell$, and width, $d$.}

\label{fig:disk_dia}
\end{figure}

One approach to estimate the induced velocity due to the migration is the analysis proposed by \citet{Houghton2019}. In this approach, shown in figure \ref{fig:disk_dia}, the vertically migrating aggregation as an actuator disk with diameter, $D$, that is in a steady upward climb (i.e., no acceleration) of velocity,  $W_{v}$. From a force balance, assuming steady climb (i.e., no acceleration). Assuming the thrust force from the vertical swimming ($F_T$) is balanced by the slight negative buoyancy of the aggregation ($F_B$) and the fluid drag on the swimmers ($F_D$), the force balance can be expressed

\begin{equation}
    |F_T| = F_B + F_D  = N  \left[\Delta \rho \frac{4 \pi a_r }{3}\left(\frac{d}{2}\right)^3 \right] + N \left[\frac{\rho}{2} C_D W_v^2  \pi\left(\frac{d}{2}\right)^2 \right]
    \label{eq:forcebalance}
\end{equation}
where, $N$ is the number of animals, $d$ is the body width of the animal, $a_r=\ell/d$ is the animal body length-to-width ratio, $\Delta \rho$ is the difference in density between the animal and the seawater. To simplify the analysis, the volume of the animal is approximated using the volume of a prolate spheroid. 

The thrust force generated by the climbing aggregation can be related to the induced velocity experienced by the aggregation, $\Delta w_0$, through the relationship
\begin{align}
    F_T &=\frac{1}{2}\rho \pi D^2 \Delta w_0 \left( W_v+\Delta w_0 \right),
    \label{eq:Ft_ind}
\end{align}
which indicates that the thrust force from the migrating aggregation is balanced by the downward momentum injected into the fluid by the swimmers. In the far field where the pressure has recovered, the thrust force from the aggregation can be related to a wake velocity, $W_w$, through
\begin{equation}
    F_T = \frac{1}{2}\rho A_D W_w^2=\frac{\pi}{8}\rho \pi D^2 W_w^2.
\end{equation}

Using equations \ref{eq:Ft_ind} and \ref{eq:forcebalance}, the induced velocity of the aggregation in climb can be solved for as 

\begin{equation}
    \Delta w_0 = \sqrt{\frac{W_v^2}{4}+\frac{(F_B + F_D)}{2 \rho A_D}} -\frac{W_v}{2} 
    \label{eq:indvel}
\end{equation}

If the disk is in a steady climb where the thrust is balanced by the weight of the aggregation and the drag on the disk, then the induced velocity term, $\frac{(F_B + F_D)}{2 \rho A_D}$, can be expressed as
\begin{align*}
\frac{(F_B + F_D)}{2 \rho A_D} &= \frac{2 N}{\rho\pi D^2} \left(\frac{\Delta\rho g 4\pi a_r}{3}\left(\frac{d}{2}\right)^3 + \frac{\rho}{2} C_D W_v^2  \pi\left(\frac{d}{2}\right)^2 \right)\\
\frac{(F_B + F_D)}{2 \rho A_D} &=\frac{d^2}{4}\frac{N}{D^2} \left(\frac{4 a_r}{3}\frac{\Delta\rho}{\rho} g d + C_D W_v^2 \right)
\end{align*}
Substituting the animal number density $\Phi  = 4N /(\pi H D^2)$ gives
\begin{equation}
\frac{(F_B + F_D)}{2 \rho A_D} =\Phi \frac{\pi H d^2}{16} \left(\frac{4 a_r}{3}\frac{\Delta\rho}{\rho} g d + C_D W_v^2 \right).
\label{eq:U_ind}
\end{equation}
Substituting equation \ref{eq:U_ind} into equation \ref{eq:indvel} gives 

\begin{equation}
    \Delta w_0 = \sqrt{\frac{W_v^2}{4}+\Phi \frac{\pi H d^2}{16} \left(\frac{4 a_r}{3}\frac{\Delta\rho}{\rho} g d + C_D W_v^2 \right)} -\frac{W_v}{2} 
    \label{eq:indvel2}
\end{equation}

In circumstances where the induced velocity of the migrating aggregation is not known or cannot be measured, equation \ref{eq:indvel2} can be employed with quantities based solely on the aggregation properties and rate of climb. 
\color{black}

\subsubsection{Linearly expanding wake model}
\label{sec:models:disc:indv}
 While the vertical velocity profile is still assumed to have a Gaussian distribution in the horizontal plane, the width of the wake is prescribed to expand linearly with the distance downstream of the aggregation following 
 \citet{Bastankhah2014,Bastankhah2016} and \citet{Shapiro2018}. In the inertial frame of the migrating aggregation climbing with an upward velocity, $W_v$, the surrounding vertical velocity field is given by:

\begin{equation}
    w(x,y,z) = - W_v - \Delta w(z)\frac{D^2}{8\varsigma_0^2}\exp \left(\frac{-(x^2+y^2)}{2\varsigma_0^2 d_w(z)^2}\right).
    \label{eq:gauss_wake}
\end{equation}

 Here the aggregation is centered on the domain origin, $\Delta w(z)$ is the vertical velocity deficit along the wake centerline, $\varsigma_0$ is the characteristic width of the wake at the streamwise location of the aggregation taken to be $\varsigma_0 = 0.235D$ \cite{Shapiro2018}, and $d_w(z)$ is the dimensionless spreading function of the wake as a function of distance downstream of the aggregation. The wake spreading function is modeled as a linear expansion similar to the Jensen wake model \cite{Jensen1983} and is given by the function
 
\begin{equation}
d_w(z) = 1+k_w \ln\left(1+\exp{\left(\frac{2(z-1)}{D}\right)}\right)
        \label{eq:jensen}
\end{equation}

\noindent from \citet{Shapiro2018} with wake expansion coefficient, $k_w \approx 0.0834$. The corresponding centerline velocity deficit for the aggregation wake is given by

\begin{equation}
    \Delta w(z) =\frac{\Delta w_0}{d_w^2(z)}\frac{1}{2}\left[1+\mathrm{erf}\left(\frac{z\sqrt{2}}{D}\right)\right]
        \label{eq:centerline_vel}
\end{equation}

 \noindent where $\Delta w_0$ denotes the induced velocity at the center of the aggregation and is associated with the thrust force from to the vertically migrating aggregation. This parameter is also assumed to depend on the properties of the aggregation and will be discussed in the following section.

A mean entertainment velocity, $u_\varrho$, due to the vertical velocity field, $w$, can be modeled for an axisymmetric mean flow field using the incompressibility condition. The continuity equation for incompressible flow in cylindrical coordinates is given by
\begin{equation}
    \nabla\cdot\mathbf{u} \implies \frac{1}{\varrho} \frac{\partial(\varrho u_\varrho)}{\partial \varrho} + \frac{\partial w}{\partial z} = - \frac{1}{\varrho}\frac{\partial u_\phi}{\partial \phi}  = 0.
    \label{eq:cyl_continuity}
\end{equation}

\noindent where $z$ is the vertical coordinate, $\phi$ be the azimuthal angle, and $\varrho$ be the radial coordinate in the horizontal. Assuming the mean velocity field to be axisymmetric lets the azimuthal velocity component, $u_\phi$ be zero and simplifies equation \ref{eq:cyl_continuity} to be
\begin{equation*}
    \frac{1}{\varrho} \frac{\partial(\varrho u_\varrho)}{\partial \varrho} = - \frac{\partial w}{\partial z}.
\end{equation*}

Solving for the radial velocity gives
\begin{equation}
    u_\varrho (z,\varrho) = - \frac{1}{\varrho}\int_0^\varrho \varrho'\frac{\partial w}{\partial z}(z,\varrho') d\varrho'.
    \label{eq:u_radial}
\end{equation}

Combining equations \ref{eq:gauss_wake}, \ref{eq:centerline_vel}, \ref{eq:jensen}, and \ref{eq:u_radial} gives an axisymmetric mean velocity field associated with the vertical migration of a low aspect ratio aggregation. This velocity field can be inserted into equations \ref{eq:greens_func_x}-\ref{eq:greens_func_z} using the following transformation into Cartesian velocities:

\begin{align*}
    u (z,\varrho,\theta) = u_\varrho \cos(\theta)\\
    v (z,\varrho,\theta) = u_\varrho \sin(\theta)\\
    w (z,\varrho,\theta) = w\\
\end{align*}

\subsubsection{Numerical Approach and Results}
\label{sec:disk:results}
Similar to the previous section, the Green's function of the wake velocity was numerically integrated over the 3D domain ranging from {$-10D$ to $10D$} in each of the $x$ and $y$ directions and from {$-400D$ to $400D$} in the vertical (i.e., z-direction) using {$[N_x,\,N_y,\,N_z] = [88,\,88,\,3520]$} grid points. The magnetic field was evaluated on a larger domain on the $yz$-plane (i.e., $x=0$) at 100 logarithmically spaced locations ranging from $x,y = 0.01 - 1000$.  As before, the Earth's magnetic field is taken to be both constant over the velocity field with a negligible declination and inclination such that only $B_y$ is nonzero. In contrast to the previous velocity models, the semi-infinite extent of the velocity signature in the vertical ($z$) direction requires that the magnetic signature now have horizontal components to ensure that the magnetic field lines are closed. These horizontal components, however, are much smaller than the vertical component due the entrainment velocity and its horizontal gradients being much smaller than those of the vertical velocity component (i.e., wake velocity). 

\begin{figure}[ht!]
    \centering
  \input{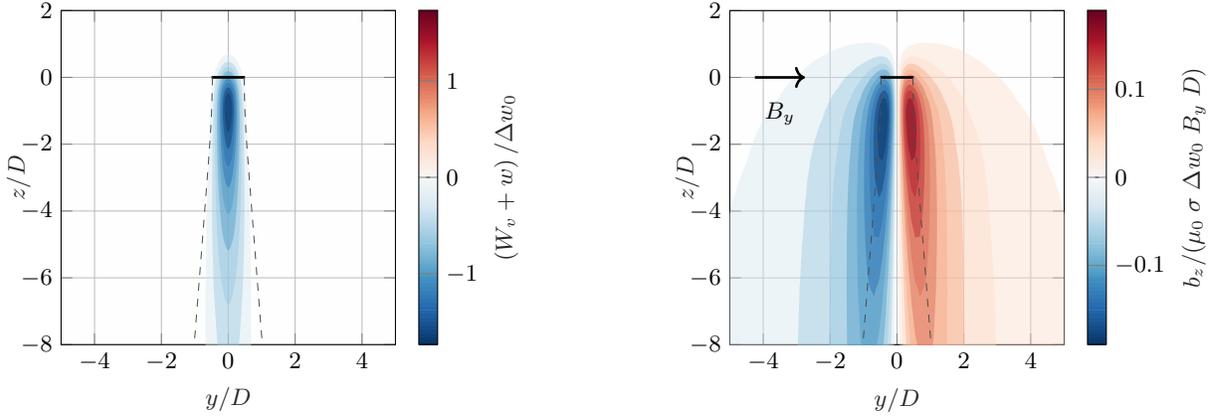}   
    \caption{Plots of the dimensionless magnetic and velocity fields. In each panel, the solid black line represents the actuator disk of diameter, $D$, and the dashed line indicates the nominal width of the aggregation wake given by $\pm2\varsigma_0d_w(z)$ in equation \ref{eq:jensen}. (a) Contour plot of the vertical wake velocity generated by the aggregation in the $yz$-plane. The velocity in the wake is largest immediately behind the aggregation and decays downstream with distance from the migration.  (b) Contour plot of the dimensionless vertical magnetic field component $(b_z/(\mu_0 \sigma \Delta w_0 B_y D))$ in the $yz$-plane. Like the wake velocity, the magnetic signature extends downstream from the aggregation and decays quickly upstream of the migration. However, as in the previous models, the magnetic signature persists further from the aggregation location along the horizontal directions than the wake velocity signature.}
    \label{fig:disk_field}
\end{figure}
Unlike the previous velocity models, which feature vertical homogeneity in the velocity field, the actuator disk model assumes a minimal velocity signature far upstream of the aggregation and an expanding jet downstream of the migration. The resulting wake velocity given by equations \ref{eq:gauss_wake} - \ref{eq:centerline_vel} is shown in figure \ref{fig:disk_field}(a) as a contour map in the $yz$-plane with the nominal wake spreading of $2\varsigma d_w(z)$ (see equation \ref{eq:jensen}) shown as a dashed line. The vertical component of the associated magnetic signature is shown in figure \ref{fig:disk_field}(b) as a contour map in the $yz$-plane against the same wake spreading function. Despite the reduced vertical extent of the wake, the magnetic signature is still observed to persist at horizontal distances much larger than the wake width for all values of $z$ downstream of the aggregation. 
 
\begin{figure}[ht]
    \centering
    \input{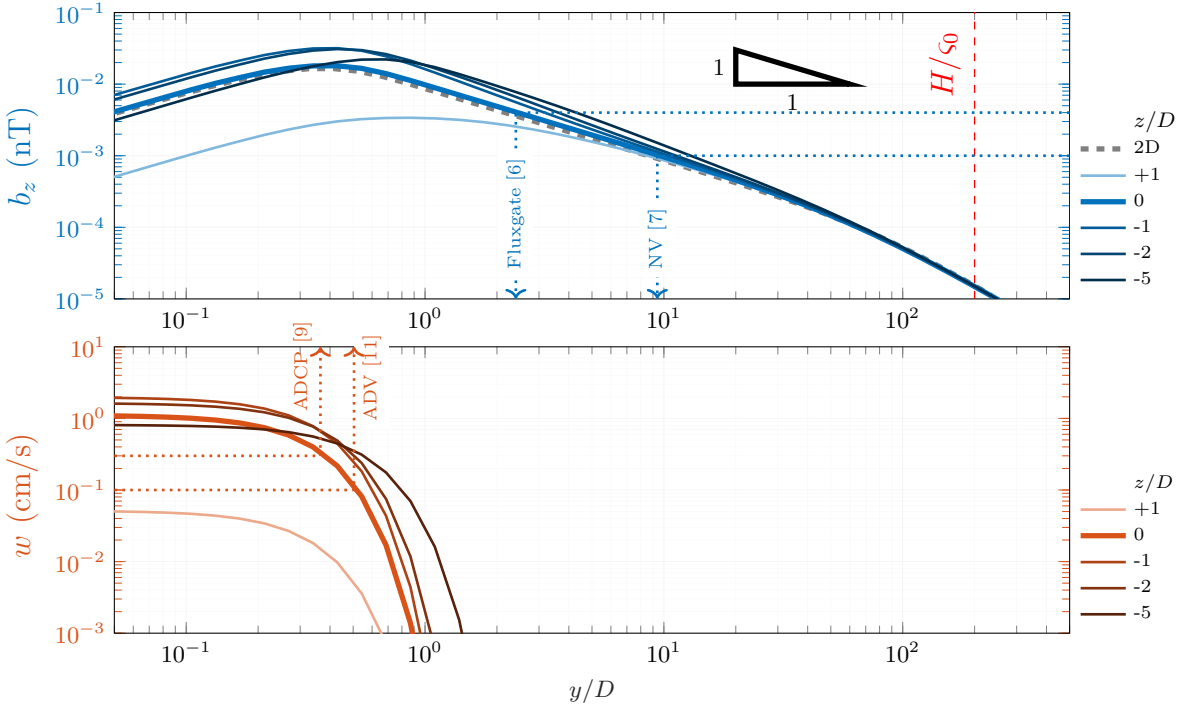}
    \caption{Profile of a representative dimensional (a) aggregation wake velocities and (b) corresponding vertical magnetic field strengths as a function of distance along the $y$-axis at different heights.  Representative values of $B_{geo}=25\mu$T, $D=100$ m, $\sigma = 5$ S/m, $\Delta w_0=1$ cm/s, $k_w=0.0834$ and $\varsigma_0=0.235D$ were chosen for the migration giving a representative magnetic signature magnitude of $157$ pT. Solid orange and blue lines show the vertical wake velocity and vertical magnetic field component along the y-axis ($x=0$) at vertical locations $z/D=-5,-2,-1,0,\;\mathrm{and}\;1$. Darker shades of each respective color indicate lower heights with the thick line denoting the height of the aggregation itself.  Dotted lines indicate the typical resolution or sensitivity limit of corresponding velocimetry and magnetometry techniques. The dashed gray line in (b) indicates the nominally equivalent magnetic signature generated a 2D Gaussian jet of infinite extent (see equation \ref{eq:bz_nd}). Red dashed line represents the height of the resolve velocity wake.}

    \label{fig:profile_disk}
\end{figure}

 Substituting representative parameters $B_{geo}=25\mu$T, $D=100$ m, $\sigma = 5$ S/m, and $\Delta w_0=1$ cm/s into these distributions again gives a magnetic signature scale of $\mu_0 \sigma \Delta w_0 B_y D= 157$ pT. Select profiles of the vertical magnetic and velocity components are shown along the $y$-axis in figure \ref{fig:profile_disk} with the respective resolution/sensitivity limits of different measurement techniques. Similar to the previous analysis, common techniques such as Acoustic Doppler Current Profilers (ACDPs) \cite{NortekVector300,NortekAquedopp6000m,Park2021} and Acoustic Doppler Velocimeters  \cite{NortekSignature1000,TeledyneWorkhorse2009,Cisewski2009,TeledyneOceanSurveyor2009,Cisewski2021} are all still suitable for observing upwelling and downwelling currents from migrating aggregates of zooplankton even up to $5D$ downstream of the aggregation. 
As can be seen in figure \ref{fig:profile_disk}, the Gaussian decay of the velocity signature still confines the usefulness of velocimetry techniques to the immediate vicinity of the wake. However, because of the gradual expansion of the wake downstream of the aggregation, the horizontal distance from the axis of the migration where the velocity signature can be detected gradually increases downstream of the migration. 
 
By comparison, the limited vertical extent of the wake in this model has somewhat reduced the overall magnitude of the magnetic signature distribution and choice of normalization using the aggregation size versus wake width. Furthermore, the power-law decay of the signature is slightly faster than the $y^{-1}$ predicted by the previous models. This feature is evidenced by the dashed gray line showing the nominal results from the previous section for comparison. However, at fixed horizontal distances, there is a relative enhancement of the magnetic signature with downstream distance from the migration outside the velocity wake due to the entrainment and wake spreading. While the detection distances for state-of-the-art fluxgate magnetometers \cite{Bartington2022,Magson2S,THM1186} and Nitrogen-vacancy (NV) centers \cite{Wolf2015} are significantly reduced compared to the previous models, each detection distance still extends approximately an order of magnitude further than the velocimetry techniques according to this model.

\section{Measurement Techniques}
\label{sec:tables}



\newlength{\tabwid}
\setlength{\tabwid}{1.5in}
\begin{turnpage}

    \begin{table}[ht!]
        \centering
         \caption{\textbf{Marine Velocimetry Techniques}}
        \label{tab:tab1_vel_params}
        \begin{ruledtabular}

        \begin{tabular}{m{\tabwid}ccccl}
    Technique &  Velocity  & Data & Velocity  & Measurement  & References\\
     &   Accuracy mm$\cdot\mathrm{s}^{-1}$ & Dimension &  Components &  Domain Size & \\
        \hline\\

    \multirow{4}{\tabwid}{Particle Image Velocimetry (PIV)}
         &  $\sim 1$     &   Area (2D)   &    $u$,$v$  &  $0.200\, \mathrm{m} \times 0.200\,\mathrm{m}$ &    \citet{Bertuccioli1999}\\    
         &  $\sim 2$     &   Area (2D)   &    $u$,$v$  &  $0.150\, \mathrm{m} \times 0.150\,\mathrm{m}$ &    \citet{Katija2008}\\    
         &  $\sim 1$     &   Area (2D)   &    $u$,$v$  & $0.075\, \mathrm{m} \times 0.095\,\mathrm{m}$ &    \citet{WangPIV2012}\\ 
          &  $\sim 1$    &   Area (2D)   &    $u$,$v$  &  $0.100\, \mathrm{m} \times 0.200\,\mathrm{m}$ &    \citet{Jin2019}\\  
        &      &       &      &      & \\
      &      &       &      &   \underline{Cyl. - $(\mathrm{dia.} \times \mathrm{ht.})$}   & \\
     \multirow{2}{\tabwid}{Acoustic Doppler Velocimeter (ADV)}     &  $1$  $\pm$ 0.5\%    &   Point   &    $u$,$v$,$w$  &  $0.015\, \mathrm{m}\, \times 0.005-0.020\, \mathrm{m}$ & Nortek Vector - 300 m \cite{NortekVector300} \\ 
        &  $5$  $\pm$ 1\%    &   Point   &    $u$,$v$,$w$  &  ~$0.75\, \mathrm{m}$ & Nortek Aquadopp - 6000 m \cite{NortekAquedopp6000m}   \\ 
        &  $30$     &   Point   &    $u$,$v$,$w$  &  ~$0.006\,\mathrm{m}\times\,0.001-0.0091\,\mathrm{m}$ & 
        \citet{Park2021}\\ 
        &      &       &      &      & \\

    \multirow{4}{\tabwid}{Acoustic Doppler Current Profiler (ADCP)}               &      &       &      &    \underline{Range (Cell Size)}   & \\         
    &  $3$ $\pm$ 0.3\%    &   Linear (1D)   &    $u$,$v$,$w$  &  $30\, \mathrm{m}\,(0.2-2\, \mathrm{m})$ & Nortek Signature - 1000 m \cite{NortekSignature1000}\\
        &  $5$  $\pm$ 1\%   &   Linear (1D)   &    $u$,$v$,$w$  &  $600\, \mathrm{m}\,(4-32\, \mathrm{m})$ & Teledyne Workhorse Long Ranger \cite{TeledyneWorkhorse2009}\\  
                &      &       &      &      & \citet{Cisewski2009}\\ 
     &  $5$  $\pm$ 1\%   &   Linear (1D)   &    $u$,$v$,$w$  &  $>400\, \mathrm{m}\,(8\, \mathrm{m})$ & Ocean Surveyor 150kHz \cite{TeledyneOceanSurveyor2009}\\   
        &      &       &      &      & ,\citet{Cisewski2021}\\ 
                &      &       &      &      & \\ 
    \multirow{3}{\tabwid}{Laser Doppler Velocimeter (LDV)}       &      &       &      &  \underline{Cyl. - $(\mathrm{dia.} \times \mathrm{ht.})$}   & \\
         &  \na    &   Point   &    $u$,$v$  &  $0.0001\, \mathrm{m}\, \times 0.00015\, \mathrm{m}$ & \citet{Trowbridge1995}  \\   
      &  $2.7$     &   Point   &    $u$,$w$  &  $0.0003\, \mathrm{m}\, \times 0.003\, \mathrm{m}$ &   \citet{Agrawal1992}\\  
    &  $\sim 0.01$     &   Point   &    $u$  &  $0.00022\, \mathrm{m}\, \times 0.0047\, \mathrm{m}$ &   \citet{Fowlis1973}\\   
        &      &       &      &      & \\


        \end{tabular}
        \end{ruledtabular}

    \end{table}
        
\end{turnpage}



\setlength{\tabwid}{2in}
\begin{turnpage}
    \begin{table}[ht!]
        \caption{\textbf{Magentometry Techniques}: }
        \label{tab:tab2_mag_params}
        \centering
                    \begin{ruledtabular}
            \begin{tabular}{m{\tabwid}ccccl}
        Technique &  Measurement  &  Sensitivity  & Resolution & Nominal  & Reference\\
             &  Type & (nT/$\sqrt{\mathrm{Hz}})$  & (nT) &  Instrument Size  & \\
           \hline 
                &        &   &       &   &    \\

        \multirow{1}{\tabwid}{Fluxgate Magnetometer} &  Vector      &  $<4\times10^{-3}$ at 1 Hz  &    \na   & $3.0 \times 3.0 \times 25.0$ cm$^3$  & Barington Instruments Mag-13 \cite{Bartington2022}     \\    
            &  Vector      &  $<15 \times 10^{-3}$ &   $7.75\times10^{-3}$    &  $5.0-7.0$ cm dia. $\times$ 4.0 cm  &   Magson GmbH MFG-2S\cite{Magson2S} \\    
        &   Vector   &    \na   &  $\pm100$   & $7.0\times3.0\times3.2\;\mathrm{cm}^3$ & Metrolab THM1186 \cite{THM1186}\\  
        &   Vector   &    $<0.7$   &  $ 0.1$   & $1.0$ cm (dia.) $\times$ $4.0$ cm  (ht.) & Fluxmaster \cite{StefanMayerInstruments2002}\\  
            &        &   &       &   &    \\
    
         \multirow{2}{\tabwid}{Superconducting Quantum Interference Device (SQUID)\footnote{Requires cryogenic cooling} }    & Vector   & $7\times 10^{-6}$ \footnote{at 77.4 K}& $10\times10^{-9}$ &\na&  \citet{Faley2006}\\   
           &        &   &       &   &    \\
                    &        &   &       &   &    \\

        \multirow{1}{\tabwid}{Nitrogen-Vacancy Center } 
            &  Vector      &  $900\times10^{-6}$  &    $100\times 10^{-6}$   &  \na  &    \citet{Wolf2015}\\    
            &  Vector      &   $7$&   \na    &  $7\times7\times11\;\mathrm{cm}^3$  &    \citet{Webb2019}\\ 
            &  Vector      &  $~9$  &    $57.6$   &  $\sim1.5$ cm (dia.) $\times$ 3 cm  (ht.)&   \citet{Kuwahata2020}\\  
                            &        &   &       &   &    \\

        \multirow{1}{\tabwid}{Inductive Pickup Coil \footnote{Not suitable for DC magnetic fields.}}       &  Vector    &   $100\times10^{-6} $at 1 Hz    &  $2 \times 10^{-6}$   & $\sim5$ cm dia. $\times$ 110 cm  & \citet{Chen2020}\\
             &   Vector  &   $75\times10^{-3} $at 0.01 Hz  &   \na   & $\sim7$ cm dia. $\times$ 134 cm & KMS Technologies LEMI-152 \cite{LEMI152}\\  
            &      &       &    &  &\\
    
        \multirow{1}{\tabwid}{Hall-Effect Sensor}  &   Vector   &    \na   &  $\pm20\times10^3$   & $7.6\times2.25\times1.4\;\mathrm{cm}^3$ & Metrolab THM1176 \cite{THM1186}\\
        &      &       &    &  &\\
    
        \multirow{1}{\tabwid}{\underline{Optically Pumped Magnetometer}}       &      &       &    &  &\\
     
         Potassium Vapor     &  Total Field    &   $200\times 10^{-6}$   &   $100\times10^{-6}$   & $\sim5$ cm dia. $\times$ 10 cm  &  GEM GSMP-35 \cite{GSMP35}\\   
    
    Rubidium-87 Vapor &   Total Field   &    $14\times10^{-6}$   & \na   &  \na &  \citet{Lucivero2021}\\
         &  Total Field    &   $14\times10^{-6}$   &  \na    &\na  & \citet{Limes2020}\\   
                   
       SERF\footnote{Spin Exchange Relaxation-Free, Requires near zero magentic field conditions} Potassium Vapor &   Vector  & $540\times10^{-9}$ & \na & $\sim0.5$ cm (dia.) $\times$ 2.1 cm (ht.) &\citet{Kominis2003}\\         
        &  Vector    &   $160\times10^{-9}$   & \na  & $10$ cm (dia.) $\times$ $10$ cm (ht.)&  \citet{Dang2010}\\   
         &        &   &       &   &    \\
         
            \end{tabular}

            \end{ruledtabular}

        \end{table}
        \end{turnpage}


\bibliographystyle{apsrev4-2}
\bibliography{refs_supp.bib}